\documentclass[amsmath,amssymb,amstext,aps,floats,amsfonts,notitlepage,superscriptaddress,eqsecnum,nofootinbib,10pt]{revtex4-1}
\usepackage{graphicx}       
\usepackage{dcolumn}        
\usepackage{bm}             
\usepackage{tensor}
\usepackage[colorlinks]{hyperref}
\usepackage{color}
\usepackage[usenames,dvipsnames,svgnames,table]{xcolor}
\usepackage{float}
\allowdisplaybreaks[1]

\usepackage{dcolumn}
	\newcolumntype{.}{D{.}{.}{13}}
	\newcolumntype{d}[1]{D{.}{.}{#1}}
\usepackage{longtable}
\usepackage{subfigure}
\usepackage{rotating}
\usepackage[small,compact]{titlesec}
\usepackage{etoolbox}
\makeatletter
\patchcmd{\ttlh@hang}{\parindent\z@}{\parindent\z@\leavevmode}{}{}
\patchcmd{\ttlh@hang}{\noindent}{}{}{}
\makeatother
\usepackage{fancyhdr}
\usepackage{times}
\usepackage{multirow}
\usepackage{sidecap}
\usepackage{eurosym}
\usepackage{setspace}

\definecolor{CiteColor}{rgb}{0,0.5,0}
\hypersetup{citecolor=CiteColor}
\definecolor{RefColor}{rgb}{0.55,0,0}
\hypersetup{linkcolor=RefColor}
\definecolor{darkgreen}{rgb}{0.2,0.7,0.2}

\newcommand{\beq}{\begin{equation}}
\newcommand{\eeq}{\end{equation}}

\newcommand{\bs}{\mathbf{s}}
\newcommand{\bS}{\mathbf{S}}
\newcommand{\bp}{\mathbf{p}}
\newcommand{\bn}{\mathbf{n}}
\newcommand{\bom}{\boldsymbol{\omega}}
\newcommand{\bOm}{\boldsymbol{\Omega}}
\newcommand{\tilE}{\tilde{E}}
\newcommand{\tilL}{\tilde{L}}
\newcommand{\nn}{\nonumber}
\newcommand{\EE}{E}
\newcommand{\LL}{L}
\newcommand{\KK}{\mathcal{K}}

\newcommand{\ord}{\mathcal{O}}

\newcommand{\f}{\frac}
\newcommand{\pd}{\partial}

\newcommand{\veps}{\varepsilon}
\newcommand{\be}{\begin{equation}}
\newcommand{\ee}{\end{equation}}
\newcommand{\ba}{\begin{eqnarray}}
\newcommand{\ea}{\end{eqnarray}}
\newcommand{\bi}{\begin{itemize}}
\newcommand{\ei}{\end{itemize}}
\newcommand{\bef}{\begin{frame}}
\newcommand{\ef}{\end{frame}}
\newcommand{\es}{\end{small}}
\newcommand{\Lim}[1]{\raisebox{0.5ex}{\scalebox{0.8}{$\displaystyle \lim_{#1}\;$}}}
\newcommand\T{\rule{0pt}{2.6ex}}       
\newcommand\B{\rule[-1.2ex]{0pt}{0pt}} 

\newcommand{\Delhat}{\hat{\delta}}
\newcommand{\vsp}{\vspace{0.2cm}}

\begin{document}

\title{Spin-orbit precession for eccentric black hole binaries at first order in the mass ratio}

\author{Sarp Akcay}
\affiliation{The Institute for Discovery, School of Mathematics \& Statistics, University College Dublin, Belfield, Dublin 4, Ireland.}

\author{David Dempsey}
\affiliation{Consortium for Fundamental Physics, School of Mathematics and Statistics,
University of Sheffield, Hicks Building, Hounsfield Road, Sheffield S3 7RH, United Kingdom.}

\author{Sam R.~Dolan}
\affiliation{Consortium for Fundamental Physics, School of Mathematics and Statistics,
University of Sheffield, Hicks Building, Hounsfield Road, Sheffield S3 7RH, United Kingdom.}

%

\date{\today}

\begin{abstract}
We consider spin-orbit (``geodetic'') precession for a compact binary in strong-field gravity. Specifically, we compute $\psi$, the ratio of the accumulated spin-precession and orbital angles over one radial period, for a spinning compact body of mass $m_1$ and spin $s_1$, with $s_1 \ll G m_1^2/c$, orbiting a non-rotating black hole. We show that $\psi$ can be computed for eccentric orbits in both the
gravitational self-force and post-Newtonian frameworks, and that the results appear to be consistent. We present a post-Newtonian expansion for $\psi$ at next-to-next-to-leading order, and a Lorenz-gauge gravitational self-force calculation for $\psi$ at first order in the mass ratio. The latter provides new numerical data in the strong-field regime to inform the Effective One-Body model of the gravitational two-body problem.
We conclude that $\psi$ complements the Detweiler redshift $z$ as a key invariant quantity characterizing eccentric orbits in the gravitational two-body problem.
\end{abstract}
\maketitle

\section{Introduction}\label{Sec:Intro}

The year 2016 will surely come to be regarded as the \emph{annus mirabilis} of gravitational wave astronomy.  First, LIGO reported on primogenial
 detections of gravitational waves (GWs): three distinctive ``chirps'' associated with the binary black hole mergers GW150914, GW151226 and, at
  marginal statistical significance, LVT151012 \cite{Abbott:2016blz, Abbott:2016nmj, TheLIGOScientific:2016pea}. These discoveries suggest that second-
  generation ground-based detectors, operating at full sensitivity, may detect as many as 1000 black hole mergers per annum \cite{Belczynski:2016obo}.
   Second, the LISA Pathfinder mission reported on test masses maintained in almost-perfect freefall, with sub-Femto-$g$ accelerations in the relevant
    frequency band \cite{Armano:2016bkm}. The path is now clear for eLISA's launch, circa 2034.

The purpose of a space-based mission such as eLISA \cite{AmaroSeoane:2012je} is to explore the low-frequency ($10^{-4}$--$1$ Hz) gravitational wave
 sky. Key sources in this band include Extreme Mass-Ratio Inspirals (EMRIs). In a typical EMRI, a compact body of mass $m_1 \sim 1$--$10^2 M_\odot$
  [in]spirals towards a supermassive black hole of mass $m_2 \sim 10^{5}$--$10^{7} M_\odot$ under the influence of radiation reaction \cite{Barack:2003fp,
   Amaro-Seoane:2014ela}. EMRI modelling poses a stiff challenge to Numerical Relativity (NR) \cite{Sperhake:2014wpa} due to the separation of scales
    implied by the mass ratio $q \equiv m_1 / m_2 \sim 10^{-3}$--$10^{-7}$ (though see e.g.~\cite{Lousto:2010ut}); and to post-Newtonian (pN) theory
     \cite{Will:2011nz, Bernard:2015njp, Damour:2016abl}, due to the strong-field nature of the orbit. 

The gravitational self-force (GSF) programme \cite{Poisson:2011nh, Barack:2009ux, Thornburg:2011qk, Pound:2015tma}, initiated two decades ago \cite
{Mino:1996nk, Quinn:1996am}, seeks to address the EMRI challenge by blending together black hole perturbation theory \cite{Regge-Wheeler, Zerilli:1970,
 Teukolsky:1972my, Chrzanowski:1975wv}, regularization methods \cite{Barack:2001gx, Detweiler:2002mi, Vega:2007mc, Dolan:2010mt, Wardell:2015ada} and
  certain asymptotic-matching, singular-perturbation and multiple-scale techniques \cite{Mino:1996nk, Pound:2007th, Hinderer:2008dm, Pound:2009sm,
   Harte:2011ku}. The programme is influenced by some deep-rooted ideas in physics, such as Dirac's approach to radiation reaction in electromagnetism
    \cite{Dirac:1938nz,DeWitt:1960fc}, the effacement principle and effective field theory \cite{Galley:2008ih,Porto:2016pyg}. 

The ultimate aim of the GSF programme is to model the orbit (and gravitational waveform) of a typical EMRI, as it evolves over several years through 
$\sim 10^5$ orbital cycles; without making ``slow-motion'' or ``weak-field'' approximations; with a final phase error of less than a radian. Fulfilment
 of the accuracy goal requires methods for computing dissipative self-force at \emph{next-to-leading-order} in the mass ratio $q$ \cite{Pound:2012prl,
  Gralla:2012prd, Detweiler:2012prd}; see Refs.~\cite{Pound:2014xva, Pound:2014koa, Pound:2015wva, vandeMeent:2015lxa, vandeMeent:2016pee} for recent
   progress in this direction.

From one perspective, the compact body $m_1$ is accelerated away from a geodesic of the background spacetime of $m_2$ by a GSF, which may be split into
 a dissipative (``radiation reaction'') and conservative piece with respect to time-reversal. The dissipative self-force at \emph{leading} order in $q$
  has been known, in effect, since the 1970s, as it may be deduced from Teukolsky fluxes \cite{Teukolsky:1972my, Johnson-McDaniel:2014lia,
   Shah:2014tka, Sago:2016xsp}. By contrast, the more subtle consequences of the conservative self-force have only been explored in the last decade. An
    appealing perspective, offered by Detweiler \& Whiting \cite{Detweiler:2002mi} and others \cite{Thorne:1984mz, Detweiler:2005kq, Gralla:2008fg,
     Harte:2011ku}, is that a (non-spinning, non-extended) compact body follows a geodesic in a (fictitious) regularly-perturbed spacetime, $g^R_{ab} =
      \bar{g}_{ab} + h_{ab}^R$, where $\bar{g}_{ab}$ is the metric of the background spacetime parametrized by $ m_2$, and $h_{ab}^R=\ord(q)$ is a certain smooth vacuum 
perturbation obtained by subtracting a ``singular-symmetric'' piece from the physical metric. Conservative GSF effects are manifest as shifts at $\ord(q)$
 in quantities defined on geodesics.

In 2008, Detweiler \cite{Detweiler:2008ft} showed that, for circular orbits, the $\ord(q)$ shift in the redshift invariant $z = (u^t)^{-1} = \frac{d\tau}
{d t}$ for circular orbits (proportional to $h^R_{uu} \equiv h_{ab}^R u^a u^b$, where $u^a$ is the geodesic tangent vector) is independent of the
 choice of gauge in GSF theory \cite{Sago:2008id}, within a helical class. As $z$ is a physical observable (in principle at least), it may be
  calculated for \emph{any} mass ratio via complementary approaches to the gravitational two-body problem. There has emerged a concordance in results
   in overlapping domains \cite{Tiec:2014lba}, between GSF theory \cite{Detweiler:2008ft, Sago:2008id, Shah:2010bi, Shah:2013uya}, post-Newtonian
    theory \cite{Detweiler:2008ft, Blanchet:2009sd, Blanchet:2010zd, Blanchet:2013txa}, and, most recently, Numerical Relativity \cite
    {Zimmerman:2016ajr}. Moreover, the redshift invariant $z$ has been found to play a leading role in the first law of binary black hole mechanics
     \cite{LeTiec:2011ab, Blanchet:2012at, Tiec:2015cxa}.

Invariants from GSF theory, such as $z$, provide strong-field $\ord(q)$ information that can be applied to calibrate and enhance the Effective One-Body (EOB) model \cite{Damour:2009sm, Barack:2010ny, Akcay:2012ea, Akcay:2015pjz}. As the EOB model generates waveforms for binaries at $q \sim 1$, this provides a conduit for GSF results to flow towards data analysis at LIGO. A cottage industry has developed in identifying and calculating invariants
 associated with conservative GSF at leading order in $q$. For circular orbits in Schwarzschild, the invariants comprise (i) the redshift invariant
  \cite{Detweiler:2008ft, Sago:2008id} (relating also to the binary's binding energy \cite{LeTiec:2011dp}), (ii) the shift in the innermost stable
   circular orbit (ISCO) \cite{Barack:2009ey}, (iii) the periapsis advance (of a mildly-eccentric orbit) \cite{Barack:2009ey, Barack:2011ed,
    LeTiec:2011bk}, (iv) the geodetic spin-precession invariant \cite{Dolan:2013roa, Bini:2014ica, Dolan:2014pja, Bini:2015mza, Shah:2015nva}, (v) tidal eigenvalues at quadrupolar 
order \cite{Dolan:2014pja, Bini:2014zxa, Bini:2015kja}, and (vi) tidal invariants at octupolar order \cite{Bini:2014zxa, Bini:2015kja, Nolan:2015vpa}.
 There has been remarkable progress in expanding these GSF invariants to very high post-Newtonian orders \cite{Bini:2015bla, Johnson-McDaniel:2015vva,
  Kavanagh:2015lva, Shah:2015nva}. 

At present, three tasks are underway. First, the task of computing GSF invariants for a spinning (Kerr) black hole. For circular orbits, (i) the
 redshift invariant \cite{Shah:2012gu, Kavanagh:2016idg} and (ii) the ISCO shift \cite{Isoyama:2014mja} have been calculated; (iii) the periapsis
  advance has been inferred from NR data \cite{Tiec:2013twa}; but the higher-order invariants (iv)--(vi) remain to be found. Second, the task of
   identifying and computing invariants for non-circular geodesics. For eccentric orbits, the redshift is defined by $z = \mathcal{T} / T$, where 
   $\mathcal{T}$ and $T$ are the proper-time and coordinate-time periods for radial motion. The redshift has been computed numerically \cite
   {Barack:2011ed, Akcay:2015pza} and also expanded in a pN series \cite{Hopper:2015icj, Bini:2015bfb, Bini:2016qtx} for eccentric orbits in
    Schwarzschild. Recently, it was found for equatorial eccentric orbits in Kerr, in \cite{vandeMeent:2015lxa} and \cite{Bini:2016dvs}, respectively.
     Yet generalized versions of the higher-order invariants (iv)--(vi) for eccentric equatorial orbits in Schwarzschild have not yet been forthcoming,
      and generic orbits in Kerr remain an untamed frontier. Third, there remains the ongoing task of transcribing new GSF results into the EOB model
       \cite{Damour:2009sm, Barack:2010ny, Akcay:2012ea, Bini:2014zxa, Akcay:2015pjz, Bini:2015bfb, Steinhoff:2016rfi}; and comparing against NR data \cite{LeTiec:2011bk,
        Zimmerman:2016ajr}.

In this article, we consider spin-orbit precession for a spinning compact body of mass $m_1$ on an eccentric orbit about a (perturbed) Schwarzschild
 black hole of mass $m_2$. We focus on the spin-precession scalar $\psi$, which was defined in the circular-orbit context in Ref.~\cite{Dolan:2013roa}
  (see also Refs.~\cite{Bini:2014ica, Dolan:2014pja, Bini:2015mza, Shah:2015nva}). The natural definition of $\psi$ for eccentric orbits is 
\beq
\psi = \frac{\Phi - \Psi}{\Phi} ,  \label{eq:psidef}
\eeq
where $\Phi$ and $\Psi$ are, respectively, the orbital angle and the spin-precession angle (with respect to a polar-type basis) accumulated in passing from periapsis to periapsis (i.e.~over one radial period). 
We shall compute $\Delta \psi$, the $\ord(m_1)$ contribution to $\psi$ at fixed frequencies, via both GSF and post-Newtonian approaches. In the pN calculation, we work with the spin-orbit Hamiltonian $H_{SO}$ through next-to-next-to-leading-order (NNLO) \cite{Damour:2007nc, Hartung:2013dza, Hartung:2011te}, neglecting spin-squared and higher contributions.

The article is organised as follows. In Sec.~\ref{sec:test-body} we review geodesic motion and spin precession of test-bodies on the Kerr spacetime. In Sec.~\ref{Sec:GSF} we extend the gravitational self-force formalism in order to compute $\Delta \psi$ for eccentric orbits. In Sec.~\ref{Sec:pN_expansion} we calculate $\Delta \psi$ through NNLO from the spin-orbit Hamiltonian $H_{SO}$. In Sec.~\ref{Sec:Results} we present our numerical results, and  confront the GSF data with the pN expansion. We conclude in Sec.~\ref{Sec:discussion} with a discussion of future calculations. 

\emph{Conventions:}  We set $G = c = 1$ and use the metric signature $+2$. Coordinate indices are denoted with Roman letters $a,b,c, \ldots$, indices
 with respect to a triad are denoted with letters $i, j, k, \ldots$, and general tetrad indices with $\alpha, \beta, \ldots$, etc., and numerals denote
  projection onto the tetrad legs. The coordinates $(t, r, \theta, \phi)$ denote general polar coordinates which, on the background Kerr spacetime,
   correspond to Boyer-Lindquist coordinates. Partial derivatives are denoted with commas, and covariant derivatives with semi-colons, i.e., $k_{a;b} \equiv \nabla_b k_a$. 
   Symmetrization and anti-symmetrization of indices is denoted with round and square brackets, $()$ and $[]$, respectively. Overdots denote 
    derivatives with respect to proper time, i.e.~$\dot{r} = \frac{dr}{d\tau}$. 

\section{Geodesics and spin precession in the test-body limit\label{sec:test-body}}

\vsp
 \subsection{Geodetic spin precession\label{subsec:geodetic}} \vspace{0.1cm}
Let us consider a gyroscope, of inconsequential mass and size, moving along a geodesic $z^a(\tau)$ in a curved spacetime. The geodesic tangent vector
 $u^a = \frac{dz^a}{d\tau}$ (a unit timelike vector, $g_{ab} u^a u^b = -1$) is parallel-transported, such that $D u^a / d\tau = 0$, where $Dv^a/d\tau
  \equiv u^b \nabla_b v^a = dv^a/d\tau + \tensor{\Gamma}{^a _{bc}} u^b v^c$. The gyroscope's spin vector $s_a$ (which is spatial, $s_a u^a = 0$) is
   parallel-transported along the geodesic, $D s_a / d\tau = 0$; hence its norm $s^2 \equiv g^{ab} s_a s_b$ is conserved. 

We may introduce a reference basis $e^a_\alpha$ (i.e.~an orthonormal tetrad) along the geodesic, such that $g_{ab} e_\alpha^a e_\beta^b = \eta_{\alpha
 \beta}$ where $\eta_{\alpha \beta} = \text{diag}[-1,1,1,1]$ and $e_0^a = u^a$. With this basis, we may recast the parallel-transport equation $D s_a /
  d\tau = 0$ in the beguiling form
\beq
\frac{d \bs}{d \tau} = \bom \times \bs ,  \label{eq:1}
\eeq
where $(\bs)_i \equiv e_i^a s_a$, $(\bom)_i \equiv -\frac{1}{2} \epsilon_{ijk} \omega^{jk}$, and
\beq 
\omega_{ij} \equiv - g_{ab} e_i^a  \frac{D e_j^b}{d \tau} = -\omega_{ji} .
\eeq

Eq.~(\ref{eq:1}) merely describes how a parallel-transported basis varies relative to a reference basis (or `body frame'). Note that $\bom$ depends
 entirely on the choice of reference basis; in particular, if $e_i^a$ is itself parallel-transported, then $\bom = \mathbf{0}$. An obvious way forward,
  then, is to choose a physically-motivated reference basis. For example, one could choose the  eigenvectors of the $3\times3$ tidal matrix $C_{ij}
   \equiv C_{abcd} e_i^a u^b e_j^c u^d$, where $C_{abcd}$ is the Weyl tensor. As $C_{ij}$ is real and symmetric, the eigenvectors define a unique
    orthogonal basis (provided that the eigenvalues are distinct). 

Let us suppose that a natural reference basis exists, and furthermore that $\bom$ is fixed in direction (typically, orthogonal to the orbital plane).
 We may then align the reference basis so that $(\bom)_1 = 0 = (\bom)_3$ and $(\bom)_2 = \omega_{13}(\tau) = e_3^a g_{ab} \frac{De^b_1}{d\tau}$. Eq.
 ~(\ref{eq:1}) has the solution
\beq
s_1 + i s_3 = S_\perp \exp\left( i \int^\tau \omega_{13}(\tau) d\tau \right) , \quad s_2 = S_\parallel ,
\eeq
where $S_\perp$ is a complex number satisfying $s^2 = |S_\perp|^2 + S_\parallel$, with $S_\parallel$ real and constant. Thus, the precession angle $\Psi$ accumulated over one radial period $T$ is given by
\beq
\Psi = \int_0^{\mathcal{T}} \omega_{13}(\tau) d\tau \label{eq:Phis}
\eeq
where $\mathcal{T}$ is the radial period with respect to proper time. 

\vsp

\subsection{Discrete and continuous isometries\label{sec:isometry}} \vspace{0.1cm}

One may wish for for a \emph{geometric} definition of precession which does not depend on a choice reference basis. In curved space, a vector $v^i
$ parallel-transported around a closed path does not, in general, return to itself; and this immediately gives a geometric quantity. But in curved \emph
{spacetime}, timelike paths are not closed (except in pathological scenarios), so this procedure is not relevant.

For circular orbits, the spacetime and geodesic admit a continuous isometry (neglecting dissipative effects). That is, there exists a helical Killing
 vector field $k^a$ for the spacetime which aligns with the tangent vector $u^a$ on the geodesic. In Refs.~\cite{Dolan:2013roa, Bini:2014ica} a natural
  precession quantity was defined directly from the helical Killing field itself. 

By contrast, for generic orbits, there is no continuous isometry. However, for eccentric orbits in the equatorial plane there is a \emph{discrete}
 isometry, associated with the periodicity of the radial motion (neglecting dissipative effects). To make this notion more precise, let us adopt the
  passive viewpoint, in which there is a single spacetime $(\mathcal{M}, g)$  
with a local region covered by two coordinate systems $x^a$ and $x^{\prime a}$, where the transformation between coordinates is sufficiently smooth that the
 usual transformation law applies, $g^{\prime}_{ab} = \frac{\partial x^{\prime c}}{\partial x^a} \frac{\partial x^{\prime d}}{\partial x^b} g_{cd}$ (in
  the transformation law it is implicit that the left-hand and right-hand sides are evaluated at the same spacetime point). The spacetime possesses an
   isometry if the metric components in the two coordinate systems are equal when evaluated at \emph{two different spacetime points which have the same
    coordinate values in the two systems}, $x^{\prime c} = x^c$. We may extend this concept to the worldline: under this coordinate transformation $z^
    {\prime a}(\tau') = \frac{\partial x^a}{\partial x^{\prime b}} z^b(\tau)$ where $\tau'$ is a function of $\tau$, we demand that $z^{\prime a}(\tau'
     = \tau) = z^a(\tau)$.

To be more concrete, for equatorial eccentric orbits there is a discrete isometry under the linear transformation $t' = t - T$, $r' = r$, $\theta' =
 \theta$, $\phi' = \phi - \Phi$ and $\tau' = \tau - \mathcal{T}$, where $T$, $\mathcal{T}$ and $\Phi$ are the coordinate time, proper time and orbital
  angle accumulated in passing from periapsis to periapsis.

How may we exploit the discrete isometry? We may restrict attention to reference bases that respect the isometry: tetrads $e^a_\alpha(\tau)$
 transforming in the standard way $e^{\prime a}_\alpha(\tau') = \frac{\partial x^a}{\partial x^{\prime b}} e^b_\alpha(\tau)$, which satisfy $e^{\prime
  a}_\alpha(\tau' = \tau) = e^a_\alpha(\tau)$. Now consider a pair of such tetrads within the isometry-respecting class, related by $\tilde{e}_1 + i
   \tilde{e}_3 = e^{i \varphi(\tau)} (e_1 + i e_3)$. It is straightforward to show, from Eq.~(\ref{eq:Phis}), that $\tilde{\Psi} = \Psi - [\varphi
   (\mathcal{T}) - \varphi(0)]$. As both tetrads respect the isometry, the last term is, at worst, a multiple of $2 \pi$. We may eliminate this term by restricting attention to those triads that rotate once in passing around the black hole (like the spherical polar basis). Then, $\Psi$ becomes
     insensitive to the choice of reference tetrad within a rather general class. 


\vsp
\subsection{Geodetic spin precession for test bodies around black holes} \vsp

\subsubsection{Generic geodesics in Kerr spacetime}\label{Sec:Kerr_geodesics} \vspace{0.1cm}

Consider the parallel transport of spin along a generic test-body geodesic with tangent vector $u^a$ on the Kerr spacetime, in Boyer-Lindquist coordinates. The Kerr spacetime admits two Killing vectors, $X_{(t)}^a$ and $X_{(\phi)}^a$, satisfying $X_{(a;b)} = 0$, and one Killing-Yano
 tensor $f_{ab}$, satisfying $f_{(ab)} = 0$ and $f_{ab;c} = f_{[a b ; c]}$. There are three constants of motion: energy $\EE=-X^a_{(t)} u_a$, azimuthal
  angular momentum $\LL=X^a_{(\phi)} u_a$, and the Carter constant $\KK =  \tensor{f}{_a ^c} f_{b c} u^a u^b + (\LL-a\EE)^2$ \cite{Carter:1968rr}, in
   addition to the particle's mass $m_1$. 

The vector $q^a \equiv \tensor{f}{^a  _b} u^b$ is parallel-propagated along the geodesic ($u^b q_{a;b} =  f_{a c ; b} u^b u^c + f_{a c} u^b \tensor{u}
{^c _{; b}} = 0$) and orthogonal to the tangent vector ($u^a q_a = u^a f_{ab} u^b = 0$, by antisymmetry) \cite{Penrose:1973um}. Furthermore, its
 magnitude is set by the Carter constant: $q^a q_a = \KK$.

Marck \cite{Marck:1983} introduced a standard tetrad $e^a_{\alpha}$, with its zeroth leg along $u^a$ and its second leg given by $e_2^a
 =  q^a / \sqrt{\KK}$. The standard tetrad is given explicitly in Eq.~(29)--(30) of Ref.~\cite{Marck:1983}. The geometric properties of the standard
  tetrad are explored in Ref.~\cite{Bini:2016iym}.  Relative to this basis, the precession frequency is
\beq
\omega_{13} = \frac{\KK^{1/2}}{r^2 + a^2 \cos^2 \theta} \left( \frac{\EE (r^2+a^2) - a \LL}{r^2 + \KK} + a \frac{\LL - a \EE \sin^2 \theta}{\KK - a^2 \cos^2 \theta} \right).
\eeq

\subsubsection{Equatorial geodesics in Kerr spacetime}\label{Sec:Equatorial_geodesics_in_Kerr} \vspace{0.1cm}
For orbits confined to the equatorial plane ($\theta = \pi/2$), the triad legs $e_1^a$, $e_2^a$ and $e_3^a$ are eigenvectors of the
 electric tidal tensor $C_{ij}$, with eigenvalues $-(2 + 3\KK/r^2)m_2/r^3$, $(1 + 3\KK/r^2)m_2/r^3$ and $m_2/r^3$, respectively. Thus, the reference basis
  has local physical significance. The second triad leg $e_2^a$ reduces to the unit vector orthogonal to the plane. In addition, $e_\alpha^a$ in
   the plane is a function of $r$ and $\dot{r}=\frac{dr}{d\tau}$ only (and the constants of motion) and so it respects the discrete isometry (Sec.~\ref
   {sec:isometry}). The precession frequency reduces to
\beq
\omega_{13} = \frac{\sqrt{\KK}}{r^2 + \KK} \left( \EE + \frac{a}{\LL - a\EE} \right) . \label{eq:omKerr}
\eeq

\subsubsection{Equatorial geodesics in Schwarzschild spacetime\label{sec:geodesics}} \vspace{0.1cm}
Hereforth, we shall assume the black hole is non-rotating ($a=0$). In standard Schwarzschild coordinates $\{t,r,\theta,\phi\}$ the line element is
 $ds^2= g_{ab} dx^a dx^b = -f dt^2 + f^{-1}dr^2 + r^2(d\theta^2 + r^2 \phi^2)$, where $f(r) = 1 - 2M/r$. The Carter constant reduces to $\KK = \LL^2$,
 and the precession frequency to $\omega_{13} = \EE\LL/(r^2 + \LL^2)$. Explicitly, the standard tetrad has the following components:
\begin{align}
e_0^a = u^a &= \left[\EE/f, \dot{r}, 0, \LL/r^2 \right] , & 
e_2^a &= \left[0, 0, 1/r, 0 \right] , \nn \\
e_1^a &= \frac{1}{f \sqrt{1 + \LL^2/r^2}} \left[\dot{r}, f \EE, 0, 0 \right] , & 
e_3^a &= \frac{1}{r \sqrt{1 + \LL^2/r^2}} \left[\EE \LL / f, \LL \dot{r}, 0, 1 + \LL^2 / r^2 \right] , \label{eq:tetrad}  
\end{align}
with $\dot{r} = u^r$ determined from the energy equation, 
\beq
\dot{r}^2 = \EE^2 - f (1 + \LL^2 / r^2). \label{eq:energy}
\eeq 

A bound eccentric geodesic may be parametrized by
\beq
r(\chi) = \frac{p\, m_2}{1 + e \cos \chi},  \label{eq:rchi}
\eeq
where $\chi$ is the relativistic anomaly, and $p$ and $e$ are the (dimensionless) semi-latus rectum and eccentricity. The dimensionless energy $E$ and angular momentum $L$ are related to $p$ and $e$ by \cite{Akcay:2015pza}
\beq
\EE = \left[\frac{(p-2-2e)(p-2+2e)}{p(p-3-e^2)} \right]^{1/2}, \quad \LL = \frac{p\, m_2}{\sqrt{p-3-e^2}} .
\eeq
Expressions for $d\tau/d\chi$, $dt/d\chi$, and $d\phi/d\chi$ are given in Eq.~(2.6), (2.7a) and (2.7b) of Ref.~\cite{Akcay:2015pza}. To these, we supplement
\beq
\frac{d \Psi}{d \chi} = \omega_{13} \frac{d\tau}{d\chi} = \sqrt{\frac{p-3-e^2}{p-6-2e\cos\chi}} \frac{\sqrt{(p-2-2e)(p-2+2e)}}{(p-2+2e\cos\chi-e^2\sin^2\chi)} .
\eeq
We may find $\mathcal{T}$, $T$, $\Phi$ and $\Psi$ by integrating over one radial period, e.g.,
$
\Psi = \int_0^{2\pi} \frac{d \Psi}{d \chi} d\chi . 
$
The orbital angle $\Phi$ is given in terms of an elliptic integral in Eq.~(2.9) of Ref.~\cite{Akcay:2015pza},
\beq
\Phi = 4 \sqrt{\frac{p}{p-6+2e}} \text{ellipK} \left( \frac{4e}{p-6+2e} \right).
\eeq
The other quantities may not be written in a similarly compact form. However, one may easily calculate these numerically, or by expanding as a series in $1/p$ as follows:
\begin{eqnarray}
\frac{\mathcal{T}}{2 \pi \upsilon} &=& \frac{1}{j_e^3} + \frac{3}{2 j_e} p^{-1} + \left( \frac{3}{2} + \frac{6}{j_e} + \frac{3}{8} j_e \right) p^{-2} +
 \left(\frac{29}{2} + \frac{24}{j_e} + 3 j_e + \frac{3}{4}j_e^2 - \frac{1}{16} j_e^3 \right) p^{-3} + \nn \\ && \hspace{4.0cm} + \left(\frac{96}{j_e} +
  \frac{1737}{16} + 18 j_e + \frac{129}{16} j_e^2 - \frac{3}{4} j_e^3 - \frac{3}{16} j_e^4 + \frac{3}{128} j_e^5 \right) p^{-4} + \ord(p^{-5}) , \\
\frac{T}{2 \pi \upsilon} &=& \frac{1}{j_e^3} + \frac{3}{j_e} p^{-1} + \left(\frac{15}{2} + \frac{6}{j_e} \right) p^{-2} + \left( \frac{75}{2} + 
\frac{30 - 6e^2}{j_e} \right) p^{-3} + \left( \frac{3729}{16} + \frac{96}{j_e} + 30 j_e - \frac{75}{16}j_e^2 \right) p^{-4} + \ldots, \\
\frac{\Phi}{2\pi} &=& 1 + 3 p^{-1} +  \left( \frac{27}{2} + \frac{3}{4}e^2 \right)p^{-2} + \left(\frac{135}{2} + \frac{45}{4} e^2 \right) p^{-3} + 
\left(\frac{2835}{8} + \frac{945}{8} e^2 + \frac{105}{64} e^4 \right) p^{-4} + \ord(p^{-5}) ,  \\
\frac{\Psi}{2\pi} &=& 1 + \frac{3}{2} p^{-1} + \left(\frac{63}{8} - \frac{3}{4} e^2\right) p^{-2} + \left( \frac{675}{16} - \frac{21}{8} e^2 - 
\frac{3}{16}e^4 \right) p^{-3} + \nn \\ && \hspace{4.4cm} + \left( \frac{29403}{128} + \frac{363}{32} e^2 - \frac{249}{64} e^4 - \frac{3}{32} e^6 \right) p^{-4} + \ord(p^{-5})   ,
\end{eqnarray}
where $\upsilon = m_2 p^{\frac{3}{2}}$ and $j_e = \sqrt{1 - e^2}$. Expansions for the redshift $z = \mathcal{T} / T$ and the spin precession $\psi = 1 - \Psi / \Phi$ follow immediately,
\begin{eqnarray}
z &=& 1 - \frac{3}{2} j_e^2 p^{-1} + j_e^3 \left(-6 + \frac{39}{8}j_e \right)  p^{-2} + j_e^3 \left(-23 + 6 j_e + 30 j_e^2 - \frac{235}{16} 
j_e^3 \right)  p^{-3} + \nn \\ && \hspace{4.4cm} + j_e^3 \left(-\frac{249}{2} + 24 j_e + 174j_e^2 + 6 j_e^3 - \frac{507}{4} j_e^4 + \frac{5643}{128}
 j_e^5 \right) p^{-4} + \ord(p^{-5}) , \\
\psi &=&\phantom{1 + \;} \frac{3}{2} p^{-1} +  \left(\frac{9}{8} + \frac{3}{2} e^2 \right) p^{-2} + \left(\frac{27}{16} + \frac{33}{4} e^2 + 
\frac{3}{16} e^4 \right) p^{-3} + \nn \\ && \hspace{4.4cm} + \left( \frac{405}{128} + \frac{705}{16}e^2 + \frac{123}{32} e^4 + \frac{3}{32} e^6  \right) p^{-4} + \ord(p^{-5}) . \label{eq:psibar}
\end{eqnarray}
Note that, in the circular limit $e \rightarrow 0$, we have the exact results $z = \sqrt{1 - 3/p}$ and $\psi = 1 - \sqrt{1 - 3/p}$. 

The radial and (average) azimuthal frequencies are defined via
\beq
\Omega_r = \frac{2 \pi}{T} , \quad \quad \Omega_\phi = \frac{\Phi}{T} . \label{eq:frequencies}
\eeq
The frequencies $\{\Omega_r, \Omega_\phi\}$ can be measured by an observer at infinity, and thus provide an unambiguous parametrization for eccentric
 orbits. By contrast, the semi-latus rectum $p$ and eccentricity $e$ are defined only in terms of the periapsis and apsis radii, which are not invariant under a
  change of radial coordinate. 

\vsp

\section{Gravitational self-force method\label{Sec:GSF}} \vsp
In Sec.~\ref{sec:test-body} we examined spin precession $\psi$ in the test-body limit. We now consider a compact body with a finite mass $m_1 \ll m_2$
 and a small spin $s_1 \ll (G / c) m_1^2$, orbiting a Schwarzschild black hole of mass $m_2$, where $m_1$ and $m_2$ are the ADM masses. \vsp

\subsection{Outline of scheme}\label{Sec:Outline_GSF_method} \vspace{0.1cm}
We start by assuming that there exists a well-defined function $\psi(m_2 \Omega_r, m_2 \Omega_\phi, q)$, for \emph{any} mass ratio $q=m_1/m_2$. We seek
 to isolate and compute the linear-in-$q$ part of this function using perturbation theory, that is,
\beq
\Delta \psi(m_2 \Omega_r, m_2 \Omega_\phi) \equiv  \left[ \psi(m_2 \Omega_r, m_2 \Omega_\phi, q) - \psi(m_2 \Omega_r, m_2 \Omega_\phi, 0) \right]_{\ord(q)},  \label{eq:Deltapsi}
\eeq
where the square paranthesis denote the $\ord(q)$ part. In this section, we shall denote test-body quantities using an overbar, i.e.~$\bar{\Phi}, \bar{\Psi}, \bar{e}_\alpha^a$, etc.

Underpinning our approach is the key result that, through $\ord(m_1)$, a small slow-spinning compact body follows a geodesic in a \emph{regular perturbed
 spacetime} $g^R_{ab}$ \cite{Detweiler:2002mi}; and, furthermore, its spin vector is parallel-transported in that same spacetime \cite{Harte:2011ku}.
  Here, we restrict  to the small-spin regime $s \ll G m_1^2 / c$ and neglect Mathisson-Papapetrou terms \cite{Mathisson:1937zz,
   Papapetrou:1951pa}. Note that $g^R_{ab}$ is not the physical metric; rather, it is obtained by subtracting a certain `symmetric singular' part from
    the physical metric, following the Detweiler--Whiting formulation \cite{Detweiler:2002mi} (details of the regularization procedure are given in
     Sec.~\ref{sec:Reg_params}). The regularly-perturbed metric can be written $g_{ab}^R = \bar{g}_{ab} + h_{ab}^R$, where $\bar{g}_{ab}$ is the background Schwarzschild
      spacetime and $h^R_{ab} = \ord(m_1)$ is the metric perturbation. Henceforth, we shall omit the superscript $R$.

In our perturbative approach, we shall compare quantities defined on a worldline $\gamma$ in the regular perturbed spacetime $(\mathcal{M}, g)$ with
 quantities on a reference worldline $\bar{\gamma}$ in a background spacetime $(\bar{\mathcal{M}}, \bar{g})$. 
In the regular-perturbed spacetime, we consider a geodesic (0) with proper time $\tau$, worldline coordinates $z^\alpha(\tau)$, and orbital parameters $p,e$. In the background spacetime, there are at least three possible choices of reference worldline: (1) an accelerated worldline with $(p,e)$ on the background spacetime with the coordinates $z^\alpha (\tau(\bar{\tau}))$, 
 (2) a geodesic with $(p_0, e_0)$ on the background which becomes (1) under the influence of gravitational self force, and (3) a geodesic with $(p, e)$ on the background. 
In each case, the orbit may be parameterized using Eq.~(\ref{eq:rchi}). Note that, for a given relativistic anomaly $\chi$, the coordinate difference
 between (2) and (3) is $\ord(m_1)$. Hence, at leading order in $m_1$ the instantaneous self-force computed on (3) is the same as on (2). Thus we may
  exclusively use (3) and dispense with (1) and (2). 

We use the symbol $\delta$ to denote the difference at $\ord(m_1)$ between a quantity on geodesic (0) and the same quantity on geodesic (3), implicitly
 making the choice $\bar{\chi} = \chi$ in the comparison, e.g.~$\delta e_\alpha^a = e_\alpha^a(\chi) - \bar{e}_\alpha^a(\bar{\chi}=\chi)$. Since
  geodesics (0) and (3) have the same orbital parameters $p, e$, this implies that we are comparing at the same coordinate radius $r$, though not the
   same $t$ and $\phi$ coordinates. We should emphasize that any such difference $\delta Y$ is \emph{not} gauge-invariant, in general.

We will also use $\delta$ to denote the variation in those quantities which are defined via orbital integrals, such as $\mathcal{T}$, $T$, $\Phi$ and 
$\Psi$. Let $Y$ denote some physical quantity defined by an integral around a geodesic, and let $y = dY/d\tau$ denote its local frequency. (For
 example, $Y \in \{\mathcal{T}, T, \Phi, \Psi\}$ and $y \in \{1, u^t, u^\phi, \omega_{13}\}$.) As in Sec.~\ref{sec:geodesics}, the background quantity 
 $\bar{Y}$ is found by integrating $\bar{y}$ from periapsis to periapsis,
\beq
\bar{Y} \equiv \int_0^{2 \pi} \bar{y}\, \frac{d \bar{\tau}}{d\chi}\, d\chi .
\eeq
The first-order variation $\delta Y$ is found via the integral
\beq
\delta Y = \int_0^{2 \pi} \left( \frac{\delta y}{\bar{y}} - \frac{\delta u^r}{\bar{u}^r} \right) \bar{y}\, \frac{d \bar{\tau}}{d\chi}\, d\chi \label{eq:delta_Y_generic}.
\eeq

Barack \& Sago \cite{Barack:2011ed} (henceforth BS2011) have shown how to apply the GSF formalism to calculate $\delta \Phi$ and $\delta T$ for
 eccentric orbits, and thus also the frequency shifts $\delta \Omega_r = \delta(2 \pi / T) = - (\delta T / T ) \Omega_r$ and $\delta \Omega_\phi =
  \delta( \Phi / T ) = \left(\delta \Phi / \bar{\Phi} - \delta T / \bar{T} \right)  \bar{\Omega}_\phi $. We will follow their approach, and extend it   to calculate $\delta \Psi$. 

Recall that we seek the $\ord(q)$ shift \emph{at fixed $\Omega_r$, $\Omega_\phi$} (equivalently, fixed $\Phi$ and $T$), denoted by $\Delta Y$. This is given by
\beq
\Delta Y = \delta Y - \frac{\partial \bar{Y}}{\partial \bar{T}} \delta T - \frac{\partial \bar{Y}}{\partial \bar{\Phi}} \delta \Phi .  \label{eq:Delfromdel}
\eeq
The latter terms may be found by applying the chain rule, i.e.,~
\beq
\frac{\partial \bar{Y}}{\partial \bar{T}} = \frac{\partial  \bar{Y}}{\partial p} \frac{\partial p}{\partial \bar{T}} + \frac{\partial  \bar{Y}}{\partial e} \frac{\partial e}{\partial \bar{T}} 
\eeq
with $\{ \frac{\partial p}{\partial \bar{T}}, \ldots \}$ obtained by inverting $\{ \frac{\partial \bar{T}}{\partial p}, \ldots \}$.

It follows from the definition that $\Delta T = \Delta \Phi = \Delta \Omega_r = \Delta \Omega_\phi = 0$. It follows from Eq.~(\ref{eq:psidef}) that
\beq
\Delta \psi = - \frac{\Delta \Psi}{\bar\Phi} . \label{eq:Delpsi}
\eeq
(And it follows from $z \equiv \mathcal{T} / T$ that $\Delta z = \Delta \mathcal{T} / \bar{T}$). In sections below we focus on calculating $\delta\Psi$ and thus $\Delta \Psi$. \vsp

\subsection{Circular orbit limit\label{sec:circlimit}} \vspace{0.1cm}
Here we pause to consider the circular-orbit limit of $\Delta \psi$. Naively, one might expect $\lim_{e \rightarrow 0} \Delta \psi$ to reduce to $\Delta \psi^{\text{circ}}$, the quantity calculated for circular orbits in Refs.~\cite{Dolan:2013roa, Bini:2014ica, Dolan:2014pja, Bini:2015mza, Shah:2015nva}. On the other hand, $\Delta \psi^{\text{circ}}$ is defined by comparing \emph{circular} orbits in the perturbed and unperturbed spacetimes with the same azimuthal frequency $\Omega_\phi$, whereas $\Delta \psi$ is defined by comparing not-necessarily-circular orbits in the perturbed and unperturbed spacetimes with the same \emph{pair} of frequencies $\Omega_\phi$ and $\Omega_r$. By fixing \emph{both} frequencies, the orbit in the perturbed spacetime will not necessarily be circular even when the background orbit is so. As conceptually-different comparisons are made, it is plausible that $\lim_{e \rightarrow 0} \Delta \psi \neq \Delta \psi^{\text{circ}}$, and so it proves.

Let us consider the definitions in the $e \rightarrow 0$ limit, 
\begin{eqnarray}
\Delta \psi^{\text{circ}} &=& \delta \psi^{\text{circ}} - \frac{d \bar{\psi}^{\text{circ}}}{d \bar{\Omega}_\phi} \delta \Omega_\phi^\text{circ} , \\
\lim_{e \to 0} \Delta \psi &=& \lim_{e \to 0} \delta \psi - \lim_{e \rightarrow 0} \left[ \frac{\partial \bar{\psi}}{\partial \bar{\Omega}_r} \delta \Omega_r +   \frac{\partial \bar{\psi}}{\partial \bar{\Omega}_\phi} \delta \Omega_\phi  \right] \label{eq:Delta_psi_e20_limit}.
\end{eqnarray}
It is straightforward to establish that the first terms are equal: $\delta \psi^{\text{circ}} = \lim_{e \rightarrow 0} \delta \psi$. On the other hand, the latter terms are not, and we are left with an offset term,
\begin{eqnarray}
\lim_{e \to 0} \Delta \psi - \Delta \psi^{\text{circ}} &=&  \frac{d \bar{\psi}^{\text{circ}}}{d \bar{\Omega}_\phi} \delta \Omega_\phi^
\text{circ}  -  \lim_{e \to 0} \left[ \frac{\partial \bar{\psi}}{\partial \bar{\Omega}_r} \delta \Omega_r +   \frac{\partial \bar{\psi}}{\partial \bar{\Omega}_\phi} \delta \Omega_\phi  \right]  \\
&=&   - \f{2\, p^{3/2}\sqrt{p-3}\, (p-6)^{3/2}}{4p^2-39p+86}\left[  \delta \Omega_r^{e\to0}-\f{p-8}{\sqrt{p(p-6)}} \,\delta \Omega_\phi^{e\to0}  \right]  \nn\\
 &=& - \frac{2 \sqrt{p-3}\, (p-6)^{5/2}}{p\,(4p^2-39p+86)} \frac{\Delta \Phi^{e\to0}}{2\pi} , \label{eq:circoffset}
\end{eqnarray}
where we have used 
\begin{align}
\lim_{e \to 0}  \frac{\partial \bar{\psi}}{\partial \bar{\Omega}_r}  &=-\f{2\,p^{3/2}\sqrt{p-3}\, (p-6)^{3/2}}{4p^2-39p+86},\\
\lim_{e \to 0}  \frac{\partial \bar{\psi}}{\partial \bar{\Omega}_\phi} &=\f{d\bar{\psi}^\text{circ}}{d\bar\Omega_\phi} +\f{2\,p(p-8)(p-6)\sqrt{p-3}}{4p^2-39p+86}, 
\end{align}
and
\begin{align}
\Delta \Phi^{e\to0} &= \delta \Phi^{e\to0} -\lim_{e\to0} \frac{d \bar{\Phi}}{d \bar{\Omega}_{\phi}} \delta \Omega_\phi = -\f{\bar{\Phi}^{e\to0}}{\bar\Omega_r^{e\to0}}\left(\delta \Omega_r^{e\to0}-\f{p-8}{\sqrt{p(p-6)}} \delta \Omega_\phi^{e\to0}  \right) \nn \\
&=-\,\f{2\pi p^{5/2}}{p-6}\left(\delta \Omega_r^{e\to0}-\f{p-8}{\sqrt{p(p-6)}} \delta \Omega_\phi^{e\to0}  \right).
\end{align}
We recognise $\Delta \Phi^{e\to0} / (2 \pi)$ as the $\ord(q)$ part of the periapsis advance per unit angle $K$ \cite{Damour:2009sm, Barack:2010ny, Barack:2011ed}. It is a physical quantity which is gauge-invariant (in the usual GSF sense), and it has been calculated elsewhere. The significance of this quantity in GSF theory, pN theory and numerical relativity was explored in Ref.~\cite{LeTiec:2011bk}. Thanks to recent advances in GSF technology, $\Lim{e\to0}\Delta\psi$ can now be computed analytically up to $\ord(p^{-19/2})$ via the pN expansion for $\Delta\psi^{\text{circ}}$ of Refs.~\cite{Shah:2015nva, Kavanagh:2015lva} known up to $\ord(p^{-37/2})$ ($\ord(p^{-49/2})$ online \cite{Wardell_web}) and the EOB $\rho$ function of Ref.~\cite{Bini:2016qtx} known up to $ \ord(p^{-19/2})$ using $\Delta \Phi^{e\to0} / (2 \pi) = -(1-6y)^{-3/2}\rho/2$, where $y\equiv 1/p$. For computations done in Lorenz gauge, which is not asymptotically flat, we must add $2qy/(1-6y)^{3/2}$ to $\Delta \Phi^{e\to0} /(2 \pi)$.

In Appendix \ref{sec:AppA} we give an explicit expression for $\Delta\psi_{e^0}^{9.5}$: the analytically-known part of $\Lim{e\to 0}\Delta\psi$ up to $\mathcal{O}(y^{19/2})$. A comparison of $\Delta\psi_{e^0}^{9.5}$ and our numerical results for $\Lim{e\to 0}\Delta\psi$ is shown in Fig.~\ref{Fig:e20_limit_plot}.

\vsp 
\subsection{Formulation}\label{Sec:Formulation_GSF_method}
For notational simplicity, we now drop the over-bar notation for denoting background quantities.  \vspace{0.1cm}

\subsubsection{Perturbation of the tetrad}\label{Sec:Perturbation_of_tetrad}
We start by writing the perturbed tetrad $e_\alpha^a$ in the following way,
\begin{align}
\delta u^a = \delta {e}_0^a &= c_{00}u^a  +c_{01} e_1^a + c_{03} e_3^a , \nn \\
\delta e_1^a &= c_{10} u^a  +c_{11} e_1^a + c_{13} e_3^a , \nn \\
\delta e_3^a &= c_{30} u^a  +c_{31} e_1^a + c_{33} e_3^a .  \label{eq:eperturbed}
\end{align}
where $c_{\alpha \beta}$ are coefficients at $\ord(m_1)$, to be deduced. (N.B.~the second leg $e_2^a$ remains orthogonal to the orbital plane.) Now we
 impose the orthogonality conditions $g_{ab} e_\alpha^a e_\beta^b = \eta_{\alpha \beta}$, to deduce that
\begin{align}
c_{00} &= \frac{1}{2} h_{00} &
c_{11} &= -\frac{1}{2} h_{11}, &
c_{33} &= -\frac{1}{2} h_{33}, \nn \\
c_{10} &= h_{01}  + c_{01}, &
c_{30} &= h_{03}  + c_{03}, &
c_{13} + c_{31} &= -h_{13} , \label{eq:cco}
\end{align}
where
\beq
h_{\alpha \beta} \equiv h_{ab} e_\alpha^a e_\beta^b .
\eeq
The tangent vector ($u^a = e_0^a$) may be written in the following form, 
\beq
u^a = \left[ f^{-1} ( E + \delta E ), \dot{r} + \delta \dot{r}, 0, (L + \delta L) / r^2 \right]. \label{eq:uperturbed}
\eeq
The quantities $\delta E$ etc.~are straightforwardly related to the quantities $\Delhat E$ etc.~used in BS2011 \cite{Barack:2011ed} (appearing there as
 $\Delta E$, etc.), via $\delta E = \frac{1}{2} h_{00} E + \Delhat E$, etc. (The difference arises because BS2011 use a tangent vector $\hat{u}^a$
  normalized on the background spacetime, whereas we prefer a tangent vector normalized on the perturbed spacetime.) Note that the quantities 
  $\Delhat E$, $\Delhat L$ and $\Delhat \dot{r}$ are functions of $\chi$, i.e., they are \emph{not} constants. A procedure for calculating 
  $\Delhat E$ and $\Delhat L$ is given in Ref.~\cite{Barack:2011ed}, and $\Delhat \dot{r}$ may be deduced from the normalization condition
\beq
E \Delhat E - \dot{r} \Delhat \dot{r} - \frac{f(r)}{r^2} L \Delhat L = 0.  \label{eq:ELrdot}
\eeq
Comparing Eq.~(\ref{eq:uperturbed}) with Eq.~(\ref{eq:eperturbed}) yields
\beq
c_{01} = \frac{E \dot{r}}{f \sqrt{1 + L^2/r^2}} \left( \frac{\Delhat \dot{r}}{\dot{r}} - \frac{\Delhat E}{E} \right) ,  \quad \quad \quad 
c_{03} = \frac{\Delhat L}{r \sqrt{1 + L^2/r^2}} .  \label{eq:c01c03}
\eeq
Inserting Eq.~(\ref{eq:c01c03}) into Eq.~(\ref{eq:eperturbed}) determines the tetrad, up to the single degree of freedom implicit in 
$c_{13} + c_{31} = -h_{13}$. This local ambiguity is to be expected, as we are free to locally rotate the reference basis in the $13$ plane. The ambiguity is removed by considering the secular change over radial period, using the discrete isometry. 

\subsubsection{Perturbation of precession frequency}\label{Sec:Perturbation_of_frequency}
Now we turn attention to the leading-order variation of the precession frequency. As $\omega_{13} = g_{ab} e_3^a \frac{De_1^b}{d\tau}$ is antisymmetric
 in its indices, we shall consider the variation of $\omega_{(13)}$ and $\omega\equiv\omega_{[13]}$ separately. The former is identically zero, and it merely provides a validation check of our implementation.
Applying the variation operator $\delta$, and using the product rule $\delta(AB) = A\delta B + B \delta A$, yields
\begin{eqnarray}
0 &=& \frac{1}{4} \omega \left(h_{33} - h_{11} \right) + g_{ab} e_{(3}^a \delta \left( D e_{1)}^b / d \tau \right) , \\
\delta \omega &=& \frac{1}{4} \omega \left(h_{33} + h_{11} \right) + g_{ab} e_{[3}^a \delta \left( D e_{1]}^b / d \tau \right) .
\end{eqnarray}
Here we have used the background identities $\frac{D u^a}{d \tau} = 0$, $\frac{D e_1^a}{d\tau} = \omega e_3^a$ and $\frac{D e_3^a}{d\tau} = -\omega
 e_1^a$, and Eq.~(\ref{eq:cco}). The last terms need more careful consideration. We obtain
\begin{eqnarray}
0 &=& \frac{1}{2} \omega \left(h_{33} - h_{11}\right) - \frac{1}{2} \frac{d h_{13}}{d\tau} + \delta \Gamma_{(31)0} , \label{eq:dwsym} \\
\delta \omega &=& \frac{1}{2} \omega h_{00} + \delta\Gamma_{[31]0} + \left( c_{01} e_1^b + c_{03} e_3^b \right) e_{a[3} \nabla_b e^a_{1]} + \frac{1}{2
}\frac{d(c_{13} - c_{31})}{d\tau} ,  \label{eq:dw} 
\end{eqnarray}
where 
\begin{eqnarray}
\delta \Gamma_{[31]0} &\equiv&  \tensor{\delta \Gamma}{^a _{cd}} e_{a[3} e_{1]}^c u^d 
 = \left(\delta \Gamma_{a b c}  - h_{ad} \tensor{\Gamma}{^d_{bc}} \right) e^a_{[3} e^b_{1]} u^c ,
\end{eqnarray}
with $\delta \Gamma_{abc} \equiv \frac{1}{2} \left( h_{ab,c} + h_{ac,b} - h_{bc,a} \right)$.

Let us make several brief observations on Eqs.~(\ref{eq:dwsym})--(\ref{eq:dw}): (i) Eq.~(\ref{eq:dwsym}) is simply an identity, as can be verified by
 noting that $\delta \Gamma_{(31)0} = \frac{1}{2} e_3^a e_1^b u^c \nabla_c h_{ab} = \frac{1}{2} u^c \nabla_c h_{13} + \frac{1}{2} \omega h_{11} - 
 \frac{1}{2} \omega h_{33}$; (ii) Eq.~(\ref{eq:dw}) is not uniquely defined in a local sense, due to the ambiguity in the last term (the freedom to
  locally rotate the tetrad), but this ambiguity is eliminated once we integrate over a radial period and impose the discrete isometry condition; (iii)
   for the case of circular orbits, Eq.~(\ref{eq:dw}) agrees with Eq.~(2.65) of Ref.~\cite{Dolan:2014pja} (noting the difference in notation used in
    Ref.~\cite{Dolan:2014pja}, $(\delta \Gamma)_{\bar{3}\bar{0}\bar{1}} = \delta \Gamma_{310} + h_{ad} \tensor{\Gamma}{^d_{bc}} e^a_{3} e^b_{1} u^c$,
     and neglecting the final term); and (iv) the penultimate term in Eq.~(\ref{eq:dw}) requires the background tetrad to be treated as a field, rather
      than just a basis on the background worldline itself. 

Let us consider point (iv) in more detail. For given $E$, $L$, the background tetrad field is a function of $r$ only, with $\dot{r}(r)$ interpreted as
 a function of $r$ determined from the energy condition, $\dot{r}^2(r) = E^2 - f(r)(1+L^2/r^2)$. The tetrad is defined everywhere within $p/(1+e)$ and
  $p/(1-e)$; it is orthonormal everywhere; and $e_0^a$ is tangent to a geodesic everywhere in this region. Explicitly,
\begin{eqnarray}
c_{01} e_1^b e_{a[3} \nabla_b e_{1]}^a &=& \frac{\omega}{f r (1 + L^2/r^2)} \left[ \left(1 + \frac{L^2}{r^2} \right)^2 - \frac{E^2 L^2}{r} \right] \label{eq:c01term}
\left( \frac{\Delhat \dot{r}}{\dot{r}} - \frac{\Delhat E}{E} \right) ,\label{eq:c01_simplified} \\
c_{03} e_3^b e_{a[3} \nabla_b e_{1]}^a &=& \omega \left( \frac{1 + 2L^2/r^2}{1 + L^2 / r^2} \right) \frac{\Delhat L}{L} \label{eq:c03_simplified} . \label{eq:c03term}
\end{eqnarray}  

To proceed, we may insert Eq.~(\ref{eq:dw}) with Eqs.~(\ref{eq:c01term})--(\ref{eq:c03term}) into the integral formula,
\beq
\delta \Psi = \int_{0}^{2\pi} \left( \frac{\delta \omega}{\omega} - \frac{\delta \dot{r}}{\dot{r}} \right) \omega \frac{d \tau}{d \chi} d \chi \label{eq:delta_Psi} .
\eeq
We then find $\Delta \Psi$ using Eq.~(\ref{eq:Delfromdel}), i.e.~$\Delta \Psi = \delta \Psi - \frac{\partial \bar{\Psi}}{\partial \bar{T}} \delta T - \frac{\partial \bar{\Psi}}{\partial \bar{\Phi}} \delta \Phi$, and insert this into Eq.~(\ref{eq:Delpsi}) to obtain $\Delta \psi$. 

\vsp
\subsection{Numerical computation}\label{Sec:Numerical_computation} \vsp
Below we describe the numerical implementation of the GSF method (\ref{sec:impdetails}) and the regularization procedure (\ref{sec:Reg_params}). \vspace{0.1cm}
 
\subsubsection{Implementation details\label{sec:impdetails}} \vspace{0.1cm}
We perform the numerical computation using the GSF code of Ref.~\cite{Akcay:2013wfa} which employs a frequency-domain approach, with the method of extended
homogeneous solutions  \cite{Barack:2008ms}, to compute the components of the regularized metric perturbation $h^R_{ab}$ in Lorenz gauge. This code has already been
used (i) to evolve EMRIs via the osculating geodesics method in Ref.~\cite{Warburton:2011fk}, (ii) to obtain a large set of eccentric-orbit data for the redshift
invariant in Ref.~\cite{Akcay:2015pza}, and (iii) to numerically determine the EOB $\bar{d}$ and $q$ potentials in Ref.~\cite{Akcay:2015pjz}. We have developed the code
to compute, in addition, the scalars $h_{11}, h_{33}, \delta\Gamma_{130}, 
\delta\Gamma_{310}$ in Eqs.~(\ref{eq:dw}, \ref{eq:dwsym}). We have also calculated the regularization parameters for these quantities which we present in
Sec.~\ref{sec:Reg_params}.

The code samples the interval $\chi\in[0,2\pi]$ at 240 evenly spaced points $\chi_i$ where it outputs $h_{00}, h_{11}, h_{33}, F_t^\text{cons}, F_\phi^\text{cons},
\delta\Gamma_{130}$ and $ \delta\Gamma_{310}$ at double floating point precision. Here, $F_{t,\phi}^\text{cons}$ are the conservative parts of the $t,
\phi$ components of the GSF. We use {\it Mathematica}'s \texttt{Interpolation} function to convert the discrete data sets into continuous functions
 suitable for numerical integration. We use two different orders of interpolation, three and six, and calculate $\Delta\psi$. The change
 in $\Delta\psi$ arising from this difference is our estimated interpolation error for $\Delta\psi$.

The interpolated data is sufficiently smooth for numerical integration. However, there are a few troublesome terms that
come from double numerical integrals that arise from the $c_{01}, c_{03}$ terms. These are functions of $\hat\delta E(\chi), \hat\delta
L(\chi), \hat\delta\dot{r}(\chi)$  each of which is an integral of the components of the GSF as given by BS2011
\be
\hat\delta E(\chi) = \hat\delta E(0)\underbrace{-\int_0^\chi d\chi'\ F_t^\text{cons}(\chi') \frac{d\tau}{d\chi'}}_{\equiv \ \mathcal{E}(\chi)},\qquad \hat\delta L(\chi) = \hat\delta L(0)+\underbrace{\int_0^\chi  d\chi'\ F_\phi^\text{cons}(\chi') \frac{d\tau}{d\chi'}}_{\equiv \ \mathcal{L}(\chi)}, \label{eq:deltaE_deltaL}
\ee
where $\hat\delta E(0), \hat\delta L(0) $ are the $\ord(m_1)$ corrections to energy and angular momentum at the periapsis as explained by
BS2011. $\hat\delta\dot{r}(\chi)$ can be obtained from Eq.~(\ref{eq:ELrdot}).
Each of these terms are in turn integrated over a radial period in in Eq.~(\ref{eq:delta_Psi}), which can be problematic. We deal with this issue by first algebraically
simplifying the coefficients of $\mathcal{E}(\chi),\mathcal{L}(\chi)$ then by replacing the numerical values of the $\chi=0,\pi$  endpoints of the integrand in
Eq.~(\ref{eq:delta_Psi}) with analytical limits. More specifically, we first rewrite $\delta\omega$ using Eqs.~(\ref{eq:c01_simplified}, \ref{eq:c03_simplified}). Then,
we use $\delta u^r = \f{1}{2}h_{00} \dot{r}+\hat\delta\dot{r}$ in Eq.~(\ref{eq:delta_Psi}) to remove the
$\f{1}{2}\omega h_{00}$ term in Eq.~(\ref{eq:dw}). Finally, we rewrite the remaining terms proportional to $\hat\delta\dot{r}$ in terms of $\hat\delta E$ and $\hat\delta L$ using Eq.~(\ref{eq:ELrdot}). In the end, we are left with a term of the form
\be
f_1(\chi)\,\f{\mathcal{E(\chi)} + \mathcal{L}(\chi)}{\sin^2\chi}+ f_2(\chi)\, \f{\mathcal{E(\pi)} + \mathcal{L}(\pi)}{4\cos^2\f{\chi}{2}} + f_3(\chi)\label{eq:delta_E_delta_L_integrand} \ ,
\ee
where $f_1, f_2,f_3$ depend only on background quantities and are regular at $\chi=0,\pi$. Although Eq.~(\ref{eq:delta_E_delta_L_integrand}) looks like it diverges at
$\chi=0, \pi$, the analysis in BS2011 has proven these endpoints to be removable singularities. Therefore, we can replace the divergent values by the analytic limits at
$\chi= 0,\pi$. The details of this procedure are provided in BS2011. Let us just mention that this replacement requires evaluating $dF_t^\text{cons}/d\chi, dF_\phi^\text{cons}/d\chi$ at $\chi=0,\pi$ which we can obtain in a straightforward manner using one-sided finite-difference derivative formulae.

The GSF code uses several approximations and truncations which we explain next. The method of extended homogeneous solutions constructs the spherical-harmonically
decomposed fields $h_{ab}^{R,lm}(t,r)$ (time-domain solutions) by summing over the Fourier modes of radial motion labeled by $n$. By construction, the $n$ sum
converges exponentially, however, we can only compute a finite number of $n$ modes. So we truncate the $n$ sum by imposing a convergence criterion based on the $C^0$
continuity of the $h_{ab}^{R,lm}(t,r)$ at $r=r(\chi)\equiv r_p$: with each $n$ mode added to the sum we consider the difference $ dh \equiv \left| [ \ h_{ab}^{R,lm}(t,r\to r_p^-)
-h_{ab}^{R,lm}(t,r\to r_p^+)\ ]/h_{ab}^{R,lm}(t,r_p) \right|$ and terminate the $n$ mode computation once $dh$ reaches some prechosen threshold value. Similarly, we also
truncate the infinite sum over the multipole modes at some $l=l_\text{max}$ since the regularized modes $h_{ab}^{R,l}$\footnote{$h_{ab}^{R,l}$ are obtained from $\sum_m Y^{lm}(\pi/2,\phi)\, h_{ab}^{R,lm}$.} scale as $l^{-2}$ so this sum converges, albeit slowly \cite{Barack:2007tm}. We
approximate the remaining contribution to the $l$ sum by constructing fits to the last five to ten numerically computed $l$ modes and choosing the best fit that minimizes
the appropriate $\chi$-square. This procedure is standard for mode-sum GSF calculations and the resulting fit is referred to as the $l$-mode tail. The
details of how to calculate it can be found in, e.g., Ref.~\cite{Akcay:2013wfa}.

At each $\{p,e\}$ we run our code to produce four data sets with $\{dh,l_\text{max}\} = \{10^{-9}, 15\},\{10^{-10}, 20\},\{5\times10^{-10}, 18\},\{5\times10^{-10},
20\}$, which yield four different values for $\Delta\psi$. Recall that the raw data is discrete in $\chi$ thus is interpolated using two different orders so in the end we end up with eight different values for each $\Delta\psi(p,e)$.
Our final result for $\Delta\psi(p,e)$ is the mean value of this set and the error we quote is the difference between the maximum and minimum values. Other errors are subdominant. 

Finally, we must deal with the fact that Lorenz gauge is not asymptotically flat, i.e. $\Lim{r\to\infty}h_{tt}^{\text{R},l=0} = -2\alpha \ne 0$, where $\alpha$ is a constant. 
This gauge `unpleasantness' can be removed by transforming the original Lorenz-gauge time coordinate by $t\rightarrow t(1+\alpha)$. This shift in $t$ manifests itself in the orbital period $T$ hence in the orbital frequencies $\Omega_r, \Omega_\phi$. Returning to Eq.~(\ref{eq:Delfromdel}) and inserting this correction yields
\be
\Delta\psi = \delta\psi -(1+\alpha)\,\f{\pd\bar{\psi}}{\pd\bar{T}}\,\delta T \,- \,(1-\alpha)\,\f{\pd\bar\psi}{\pd\bar\Phi}\,\delta \Phi. \label{eq:flat_fixing}
\ee
This `flat-fixing' or `flattening-out' has become standard in Lorenz-gauge GSF computations and must be done regardless of the type of motion. Indeed,
this correction was first done for circular orbits \cite{Sago:2008id}. The details of this correction for eccentric orbits can be found in Sec.~III.B of BS2011. $\alpha$ can be computed to arbitrary precision from the monopole solution which is obtained analytically in the frequency domain 
\cite{Akcay:2013wfa}. Without this correction, one can not obtain the correct result for any invariant that one computes in Lorenz gauge.

We present our numerical results for $\Delta\psi$ in Table~\ref{tbl:data} for eccentric orbits with $p = \left\{10, 15, 20, 25, \ldots, 95, 100 \right\}$ and $ e = \{0.05, 0.075, 0.1, \ldots, 0.225, 025 \}$. We include our code's circular-orbit result, $\Delta\psi^\text{circ}$, at the top row of
 each sub-table. Recall that $\Delta\psi^\text{circ} \ne \Lim{e\to 0}\Delta\psi $ hence the sign disagreement between the $e=0$ and $e>0$ values. For each
 value $\Delta\psi(p,e)$ we display the leading digit of the corresponding error in parantheses. For example, our result for $\Delta\psi(p=10,e=0)$
 should be read as $ 5.9385659  \times 10^{-3} \pm 5\times10^{-10}$. We present analysis and plots in Sec.~\ref{Sec:Results}
\vsp

\subsubsection{Regularization}\label{sec:Reg_params} \vspace{0.1cm}
We employ the standard method of mode-sum regularization. Schematically, an unregularized quantity $X^\text{full}$ with 
$\ell$-modes $X^{\text{full}}_{\ell \pm}$ in the limit $r \rightarrow r_0^{\pm}$ is converted to a regularized quantity $X^R$ using
\beq
X^R = \sum_{\ell = 0}^\infty \left[ X^{\text{full}}_{\ell \pm} - (2l+1)A^{\pm} - B \right] \label{eq:Reg_eq_general}.
\eeq
The regularization parameters $A^\pm$ and $B$ for the relevant quantities in Eq.~(\ref{eq:dw}) are listed in Table \ref{tbl:reg}. We note that all $A^\pm$ coefficients vanish. 

\begin{figure}[h]
 \includegraphics[width=0.95\textwidth]{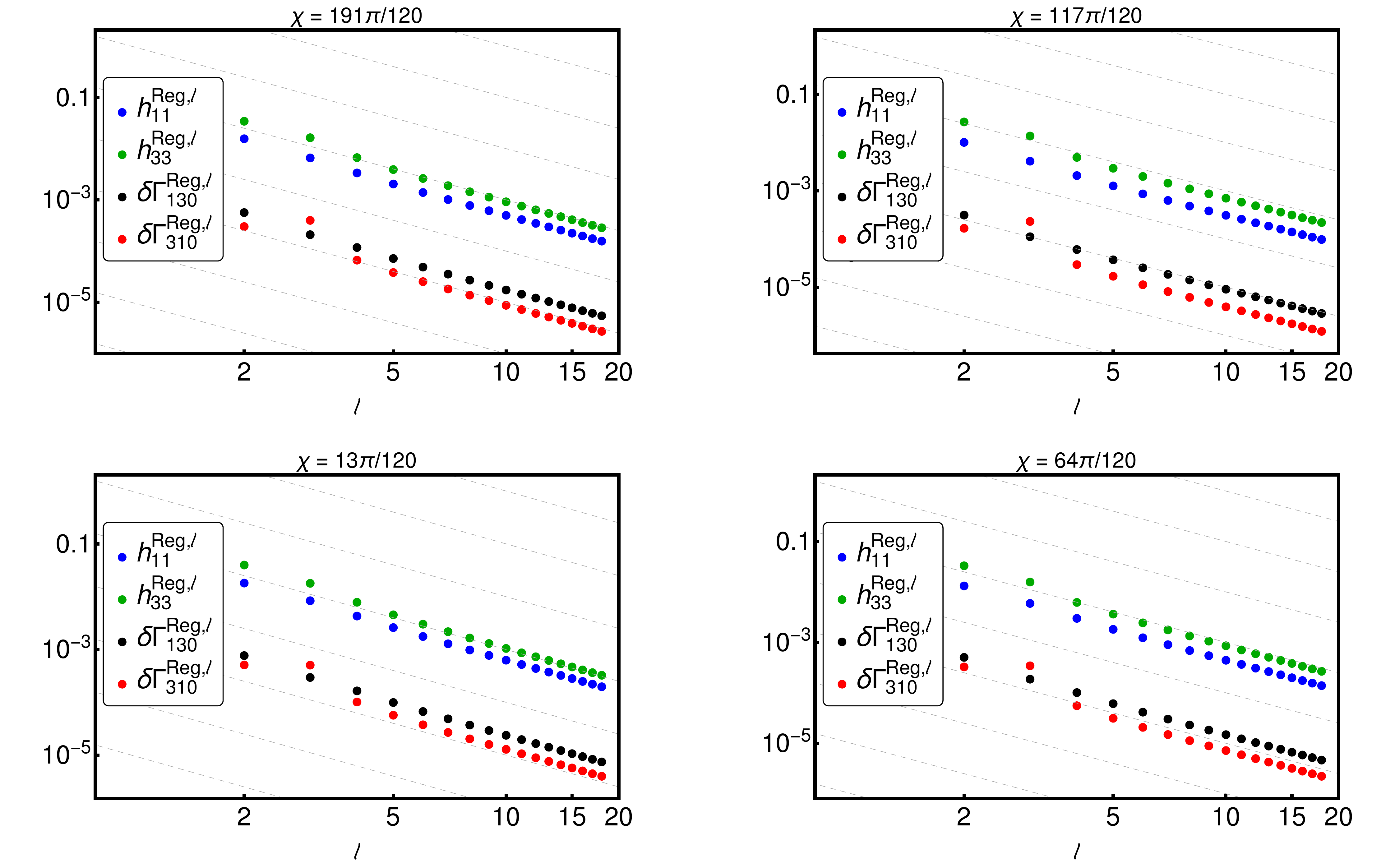}
 \caption{Log-log plots showing the regularized $l$ modes of $\left\{h_{11},h_{33},\delta\Gamma_{130},\delta\Gamma_{310} \right\} $ at four
 randomly chosen $\chi$ values along an eccentric orbit with $p=15$ and $e=0.15$ and $l_\text{max}=18$. The diagonal grid lines are $l^{-2}$ curves 
 consistent with Eq.~(\ref{eq:Reg_eq_general}).}
 \label{Fig:reg_params_plot}
\end{figure}

\begin{table}[h!]
\begin{tabular}{c|cc}
\hline
 & $A^{\pm}$ & $B$ \\
\hline
$h_{00}$, $h_{11}$, $h_{33}$ &
$0$ &
$\frac{4}{\pi \sqrt{r^2 + L^2}} \text{ellipK} \left( \frac{L^2}{r^2 + L^2} \right)$ \\
$h_{01}$, $h_{03}$, $h_{13}$ & $0$ & $0$ \\ 
$\delta \Gamma_{[31]0}$ & 0 & $\frac{4 E}{\pi L \sqrt{r^2 + L^2}} \left[ \text{ellipE} \left( \frac{L^2}{r^2 + L^2} \right)   - \text{ellipK} \left( \frac{L^2}{r^2 + L^2} \right) \right]$ . \\
\hline
\end{tabular}
\caption{Regularization parameters for metric components and first derivatives.}
\label{tbl:reg}
\end{table}
To obtain the regularization parameters (RPs) for each quantity in Table \ref{tbl:reg}, we started with RPs for the spacetime components of metric
 perturbation and its partial derivatives with respect to Schwarzschild coordinates. These were calculated by B.~Wardell \cite{Wardell:pvt} using the approach developed in Refs.~\cite{Barack:1999wf, Wardell:2009un, Heffernan:2012su, Warburton:2013lea}. We label these RPs by $A_{ab}=0,B_{ab}, A_{abc}, B_{abc}$ consistent with Eq.~(\ref{eq:Reg_eq_general}), where the rank-3 RPs are
 for $\pd_ch_{ab}$. Using the background tetrad we construct the appropriate RPs for $h_{11}, h_{33}$ etc. which we label by $B_{11}, B_{33}$ etc.
We find that $A_{130}=A_{310}=0$, which is noteworthy since the $\delta\Gamma$ terms constituted sums of up to 15 different terms for $\pd_c h_{ab} $. 
This was further confirmed by our numerical data for the unregularized $l$ modes for $\delta\Gamma_{130,310}$ which approached constant values as 
$l\to \infty$. If $A\ne0$ then we would have observed linear-in-$l$ growth for these unregularized modes. In Fig.~\ref{Fig:reg_params_plot} we show
 the regularized $l$ modes of $\left\{h_{11},h_{33},\delta\Gamma_{130},\delta\Gamma_{310} \right\} $ at four randomly chosen $\chi$ values along an
 eccentric orbit with $p=15,e=0.15$ and $l_\text{max}=18$. As expected, the regularized $l$ modes display an $l^{-2}$ powerlaw for all cases that we
 present.

\section{Post-Newtonian expansion}\label{Sec:pN_expansion} \vsp

In this section we arrive at the key result that, at next-to-next-to-leading-order (NNLO), the post-Newtonian expansion of the spin precession scalar through $\ord(q)$, $\psi = \bar{\psi} + \Delta \psi + \ord(q^2)$, is given by
\begin{align}
\bar{\psi} &= \f{3}{2} p^{-1} + \left(\f{9}{8}+\f{3}{2}e^2\right) p^{-2} + \left(\f{27}{16}+\f{33}{4}e^2+\f{3}{16}e^4\right) p^{-3} + \ord(p^{-4})\label{eq:psi_bar_p_e}, \\
q^{-1}  \Delta{\psi} &=\underbrace{- p^{-1}}_{\Delta\psi^{LO}}+\underbrace{\left(\f{9}{4}+e^2\right) p^{-2}}_{\Delta\psi^{NLO}} +\underbrace{\left[\f{739}{16}-\f{123\pi^2}{64}+\left(\f{341}{16}-\f{123\pi^2}{256}\right)e^2-\f{e^4}{2}\right] p^{-3}}_{\Delta\psi^{NNLO}}\, + \,\ord(p^{-4}) \label{eq:Delta_psi_p_e} \ .
\end{align}
In Sec.~\ref{Sec:Results} we verify that the post-Newtonian result appears to be consistent with the GSF calculation of Sec.~\ref{Sec:GSF}.

The spin-orbit (SO) Hamiltonian in ADM coordinates was presented in Ref.~\cite{Damour:2007nc} at NLO and extended to NNLO in Refs.~\cite{Hartung:2013dza, Hartung:2011te}. The Hamiltonian takes the simple form
$
H_{SO} = \sum_{i=1}^2 \bOm_i \cdot \bS_i
$
where $\bOm_i$ are spin-precession frequencies with respect to coordinate time. For equatorial orbits, 
$
\bOm_1 = \frac{\partial H_{SO}}{\partial \bS_1} = \Omega_S \mathbf{k}
$
where $\mathbf{k}$ is a unit vector orthogonal to the equatorial plane, and 
$
\Omega_S =\Omega_{S2} + \Omega_{S4} + \Omega_{S6} + \ord(c^{-8}). 
$
Here $\Omega_{S2}$, $\Omega_{S4}$ and $\Omega_{S6}$ are the contributions at LO ($c^{-2}$), NLO ($c^{-4}$), and NNLO ($c^{-6}$), respectively. 

The spin precession invariant is defined as
\beq
\psi = \frac{\left< \Omega_S \right>}{\left< \Omega_\phi \right>} , \label{eq:psiPNdef}
\eeq
where $\left< \cdot \right>$ denotes the orbital average over one radial period. (The difference between Eq.~(\ref{eq:psiPNdef}) and Eq.~(\ref{eq:psidef}) is simply due to defining precession with respect to a Cartesian-type basis as opposed to a polar-type basis). For circular orbits, $\psi$ has previously been calculated through NNLO in Ref.~\cite{Bohe:2012mr}, Eq.~(4.5) and Ref.~\cite{Dolan:2013roa}, Eq.~(9).

Our calculation is based on the approach in Sec.~IV of Ref.~\cite{Akcay:2015pza}. We perform the calculation in the centre-of-mass frame. The orbital average is taken using the generalized quasi-Keplerian (QK) representation introduced in \cite{DaDe.85}, which is known up to 3pN in both harmonic and ADM coordinates \cite{Me.al2.04}, and which is described in Sec.~IVC of \cite{Akcay:2015pza}. 

To illustrate the procedure, let us consider only the leading-order term, 
$
\bOm_{S2} = \frac{G}{c^2 r^2} \left( \frac{3 m_2}{2 m_1} \bn_{12} \times \bp_1 - 2 \bn_{12} \times \bp_2 \right)
$. 
In the centre of mass frame we have $\bp_1 = - \bp_2 = \bp$ and $\bn_{12} \times \bp_1 = \frac{L}{r} \mathbf{k}$, where $L$ is the angular momentum. Thus, $\Omega_{S2} =  \frac{g GL}{c^2 r^3}$ where $g = (3 + 3\Delta + 2\nu) / 4 \nu$ is the gyro(gravito)magnetic ratio, with $\nu \equiv m_1m_2/m^2$ the symmetric mass ratio, $\Delta \equiv (m_2 - m_1)/m$ the reduced mass difference, and $m \equiv m_1+m_2$ the total mass.

Our task is to compute the orbital averages using the QK representation. The mean anomaly $\ell = \Omega ( t - t_{\text{per}} )$ and the eccentric anomaly $u$ give the parametrization $r(u) = a_r (1 - e_r \cos u)$ and $\ell(u) = u - e_t \sin u + f_t \sin V + g_t (V - u)$, through NNLO. Here, $\Omega_r$, $a_r$, $e_t$, $e_r$, $f_t$ and $g_t$ are QK orbital elements, and $V(u)$ is specified in Eq.~(4.20) of \cite{Akcay:2015pza}. The QK representation is only complete once the orbital elements are specified in terms of orbital integrals. Following \cite{Arun:2007sg} we use the dimensionless coordinate-invariant quantities
\beq
\veps \equiv -\frac{2 \tilE}{c^2} , \quad \quad \quad j \equiv - \frac{2 \tilE \tilL^2}{(Gm)^2},   \label{eq:epsj}
\eeq
where $\tilE = E / \mu$ and $\tilL = L / \mu$ are the binding energy and orbital angular momentum, respectively, per reduced mass $\mu \equiv m_1 m_2 / m = \nu m$. Noting that $\veps = \ord(c^{-2})$ and $j = \ord(c^0)$ allows one to keep track of pN orders. The various QK elements are expanded in powers of $\veps$ in Eqs.~(4.22) of Ref.~\cite{Akcay:2015pza}. 

By way of illustration, let us consider $\left< \Omega_{S2} \right> = gGLc^{-2} \left< r^{-3} \right>$. At  NLO, we may neglect $f_t$ and $g_t$ which scale as $e_t \sim f_t \sim \ord(\veps^2)$; thus
\begin{eqnarray}
\left< r^{-3} \right> &=& \frac{\Omega_r}{2 \pi} \int_0^{2\pi} \frac{1}{r(u)^3} \frac{dt}{d \ell}\frac{d\ell}{du} du , \\
\Rightarrow \left< r^{-3} \right>^{NLO} &=& \frac{1}{2\pi a_r^3} \int_0^{2\pi} \frac{1 - e_t \cos u}{(1-e_r \cos u)^3} du = 
\frac{2 + e_r^2 - 3 e_r e_t}{2 a_r^3 (1-e_r^2)^{5/2}} .
\end{eqnarray}
To extend to NNLO, we must include the $f_t$ and $g_t$ terms; the resulting integrals are straightforward with the help of Mathematica. Other orbital averages such as $\left< r^{-4} \right>$, $\left< r^{-5} \right>$ and $\left< p_r^2 r^{-3} \right>$ etc.~may be found in a similar way.

We are led to an expression for the spin precession scalar which is valid for \emph{any} mass ratio. We may write $\psi = \psi^{LO}_\text{inv} + \psi^{NLO}_\text{inv} + \psi^{NNLO}_\text{inv}$, with
\begin{eqnarray}
\psi^{LO}_\text{inv} &=&\frac{\veps}{j} \left( \frac{3}{4} + \frac{3}{4} \Delta + \frac{\nu}{2} \right),  \\
\psi^{NLO}_\text{inv} &=& \frac{\veps^2}{j^2} \left\{ \left( \f{3}{4} + \f{3}{4}\Delta +\f{\nu}{2}\right)\left[\frac{ (48 -15\nu-18 j +  13\nu j)}{4 }\right]
+ \frac{(j-3)  \left[5 \Delta ^2+2 \Delta  (3 \nu +5)-8 \nu ^2+6 \nu +5\right]}{4 (\Delta +1)}
\right. \nn \\
&& \quad \quad + \frac{(j-1)  (\Delta -2 \nu +1) \left[5 \Delta ^2+\Delta  (10-44 \nu )-36 \nu ^2-44 \nu +5\right]}{32 (\Delta +1)^2} \nn \\
&& \quad \quad \left. + \frac{(\Delta +1)^2 (3 j-5)  \left[5 \Delta ^2+\Delta  (10-32 \nu )-12 \nu ^2-32 \nu +5\right]}{128 (\Delta -2\nu +1)} \right\} .
\end{eqnarray}
Te subscript ``inv'' is included to distinguish $\Delta\psi_\text{inv}(\veps,j)$ from $\Delta\psi(p,e)$.
The term $\psi^{NNLO}_\text{inv} = \ord(\veps^3)$ term is lengthy and will be presented elsewhere. To obtain the NNLO result, one needs an appropriate expression for the radial momentum $p_r$. Starting with Eq.~(5.6) of Ref.~\cite{Blanchet:2002mb} and taking derivatives with respect to the components of the relative velocity \cite{LeTiec:pvt}, one gets
\beq
p_r^2 = \mu^2 \dot{r}^2 \left[ 1 + c^{-2} \left\{ (1-3\nu)(\dot{r}^2 + r^2 \dot{\phi}^2) + \frac{Gm}{r} \left(6 + 4\nu \right) \right\} + \ord(c^{-4}) \right] ,
\eeq
where the overdot denotes differentiation with respect to coordinate time. 

The next step is to decompose $\psi$ into $\ord(q^0)$ and $\ord(q^1)$ parts, for comparison with the GSF result. First, we note that $\veps$ and $j$ are defined in Eq.~(\ref{eq:epsj}) in terms of energy and angular momentum, rather than the frequencies $\Omega_r$ and $\Omega_\phi$ used in the GSF approach. Following \cite{Arun:2007sg}, we introduce the dimensionless coordinate-invariant parameters,
\beq
x = \left(\frac{G m \Omega_\phi}{c^3} \right)^{2/3} , \quad \quad \quad \iota \equiv \frac{3 x}{k} ,
\eeq
where $k = K - 1$ with $K = \Omega_\phi / \Omega_r$. We replace $(\veps, j)$ with $(x, \iota)$ using Eqs.~(4.40) in Ref.~\cite{Akcay:2015pza}. Next, following \cite{Akcay:2015pza}, we introduce a pair of parameters ($y,\lambda$) better suited to the extreme mass-ratio limit $q \ll 1$, defined as $y \equiv (G m_2 \Omega_\phi / c^3)^{2/3}$ and $\lambda = 3 y /k$, so that $x = (1 + q)^{2/3} y$ and $\iota = (1 + q)^{2/3} \lambda$. After this replacement we expand $\psi$ as a series in $q$, to isolate the $\ord(q^0)$ and $\ord(q^1)$ parts. That is, we write $\bar{\psi}$ and $\Delta \psi$ as functions of $(y, \lambda)$. Finally, we switch to orbital elements $(p,e)$ which are defined with respect to the frequencies $\Omega_\phi$ and $\Omega_r$ using the functional relationships on the background spacetime. As these relationships cannot be inverted analytically, we make use of the series expansions (B1) in \cite{Akcay:2015pza}. 

At the end of this process we reach Eqs.~(\ref{eq:psi_bar_p_e}) and (\ref{eq:Delta_psi_p_e}): relative 2pN expansions for $\bar{\psi}$ and $\Delta \psi$. It is also straightforward to find higher-order-in-$q$ contributions, $\Delta \psi_{q^2}$ etc., if required for validation of any (future) second-order GSF calculation. 

The pN series pass three consistency checks. First, the pN series for $\bar{\psi}$, Eq.~(\ref{eq:psi_bar_p_e}), is in accord with the series expansion of $\psi$ on the background spacetime, Eq.~(\ref{eq:psibar}). Second, in the circular limit $e \rightarrow 0$, the difference between $\Delta \psi$ in Eq.~(\ref{eq:Delta_psi_p_e}) and $\Delta \psi^{\text{circ}}$ given by the $\mu / M$ part of Eq.~(10) in Ref.~\cite{Dolan:2013roa} is found to be precisely the pN series for the offset term in Eq.~(\ref{eq:circoffset}). Finally, the pN series (\ref{eq:Delta_psi_p_e}) appears to be consistent with the GSF numerical results for $\Delta \psi$, as we now show. 

\vsp
\section{Results}\label{Sec:Results} \vspace{0.1cm}
Here we present a selection of numerical results from the GSF method (Sec.~\ref{Sec:GSF}) and compare with the post-Newtonian series (Sec.~\ref{Sec:pN_expansion}). \vsp

\subsection{Numerical data for \texorpdfstring{$\Delta\psi$}{}}\label{Sec:Numerical_data} \vsp
Sample GSF data for $\Delta \psi$ is given in Table \ref{tbl:data}, for orbital parameters $10 \le p \le 100$ and $0 \le e \le 0.25$. 
\newpage
\begin{widetext}
\setlength{\LTcapwidth}{0.9\linewidth}
\squeezetable
\begin{longtable*}[H]{ld{17}d{17}d{17}d{17}d{17}}
176/1200 & 6.818182 & 4.244142(3)  \times 10^{-1}  &  3.108  \times 10^{-7}  &2.80(9)  \times 10^{-1}  &  9.216  \times 10^{-3} \kill 
 \caption{Numerical data for $\Delta\psi$. The first column labels the eccentricity while the remaining columns label different $p$ values. In the rows corresponding to $e=0.000$ we present the circular-orbit value, $\Delta\psi^\text{circ} $, which can be compared with Table III of Ref.~\cite{Dolan:2014pja} or with $\ord(p^{-37/2}) $ expressions of Ref.~\cite{Shah:2015nva}. Note that $\Delta\psi^\text{circ} \ne \lim_{e\to 0}\Delta\psi$ (see Sec.~\ref{sec:circlimit}). The parenthetical digits correspond to the error estimates on the last quoted significant figure. For example, $5.9385659(5)  \times 10^{-3}  = 5.9385659  \times 10^{-3} \pm 5\times10^{-10}$. \label{table:Numerical_Data} }\\
 \hline\hline
 \multicolumn{1}{l}{$e$} & \multicolumn{1}{c}{$p =10\quad$} & \multicolumn{1}{c}{$p=15$} & \multicolumn{1}{c}{$p=20$}  & \multicolumn{1}{c}{$p=25$}  & \multicolumn{1}{c}{$p=30$} \T\B \\
 \hline
 \endfirsthead
 \caption[]{(continued)}\\
 \hline\vspace{8pt}
 \endhead
 \botrule
 \endfoot
 \botrule
 \endlastfoot
0.000 & 5.9385659(5)  \times 10^{-3} &  3.3750158(2)  \times 10^{-3} &  2.07150085(7)  \times 10^{-3} &  1.38686308(2)  \times 10^{-3} &  9.9003322(2)  \times 10^{-4} \T \B  \\ 
0.050 & -5.061140(3)  \times 10^{-2} &  -4.914075(5)  \times 10^{-2} &  -4.123078(1)  \times 10^{-2} &  -3.4789491(6)  \times 10^{-2} &  -2.989822(2)  \times 10^{-2} \T \B  \\ 
0.075 & -5.052269(7)  \times 10^{-2} &  -4.911220(2)  \times 10^{-2} &  -4.1217013(8)  \times 10^{-2} &  -3.4781485(9)  \times 10^{-2} &  -2.989305(3)  \times 10^{-2}\T \B  \\ 
0.100 & -5.039852(2)  \times 10^{-2} &  -4.907223(7)  \times 10^{-2} &  -4.119775(5)  \times 10^{-2} &  -3.477028(1)  \times 10^{-2} &  -2.988568(7)  \times 10^{-2} \T \B  \\ 
0.125 & -5.023891(2)  \times 10^{-2} &  -4.902089(2)  \times 10^{-2} &  -4.1172994(7)  \times 10^{-2} &  -3.475587(2)  \times 10^{-2} &  -2.987639(1)  \times 10^{-2}\T \B  \\ 
0.150 & -5.00439(5)  \times 10^{-2} &  -4.895813(2)  \times 10^{-2} &  -4.114274(3)  \times 10^{-2} &  -3.473827(2)  \times 10^{-2} &  -2.986492(2)  \times 10^{-2} \T \B  \\ 
0.175 & -4.98131(5)  \times 10^{-2} &  -4.888402(2)  \times 10^{-2} &  -4.1107001(2)  \times 10^{-2} &  -3.471747(2)  \times 10^{-2} &  -2.9851407(5)  \times 10^{-2} \T \B  \\ 
0.200 & -4.95475(2)  \times 10^{-2} &  -4.879857(9)  \times 10^{-2} &  -4.106580(2)  \times 10^{-2} &  -3.469348(1)  \times 10^{-2} &  -2.983580(3)  \times 10^{-2} \T \B  \\ 
0.225 & -4.92468(5)  \times 10^{-2} &  -4.870169(8)  \times 10^{-2} &  -4.101909(1)  \times 10^{-2} &  -3.466629(1)  \times 10^{-2} &  -2.981813(3)  \times 10^{-2} \T \B  \\ 
0.250 & -4.89105(9)  \times 10^{-2} &  -4.859358(1)  \times 10^{-2} &  -4.096689(7)  \times 10^{-2} &  -3.463592(2)  \times 10^{-2} &  -2.979840(1)  \times 10^{-2}\T \B  \\ 
 \hline\hline
 \multicolumn{1}{l}{$e$} &\multicolumn{1}{c}{$p=35$}& \multicolumn{1}{c}{$p =40\quad$} & \multicolumn{1}{c}{$p=45$} & \multicolumn{1}{c}{$p=50$}  & \multicolumn{1}{c}{$p=55$} \T\B \\ 
 \hline
0.000 &  7.4105684(1)  \times 10^{-4} & 5.7505234(1)  \times 10^{-4} &  4.5900246(1)  \times 10^{-4} &  3.7475920(1)  \times 10^{-4} &  3.11703594(6)  \times 10^{-4} \T \B  \\ 
0.050 &  -2.6143411(7)  \times 10^{-2} & -2.319599(1)  \times 10^{-2} &  -2.083062(1)  \times 10^{-2} &  -1.889476(3)  \times 10^{-2} &  -1.728324(2)  \times 10^{-2} \T \B  \\ 
0.075 &  -2.613978(1)  \times 10^{-2}& -2.3193296(3)  \times 10^{-2} &  -2.082857(2)  \times 10^{-2} &  -1.8893114(1)  \times 10^{-2} &  -1.728196(2)  \times 10^{-2} \T \B  \\ 
0.100 &  -2.613468(2)  \times 10^{-2}& -2.318953(2)  \times 10^{-2} &  -2.0825685(1)  \times 10^{-2} &  -1.889085(5)  \times 10^{-2} &  -1.728011(1)  \times 10^{-2} \T \B  \\ 
0.125 &  -2.6128112(8)  \times 10^{-2}& -2.3184703(2)  \times 10^{-2} &  -2.082198(1)  \times 10^{-2} &  -1.888792(1)  \times 10^{-2} &  -1.727773(2)  \times 10^{-2} \T \B  \\ 
0.150 &  -2.6120105(4)  \times 10^{-2}& -2.3178794(6)  \times 10^{-2} &  -2.081744(3)  \times 10^{-2} &  -1.888434(2)  \times 10^{-2} &  -1.727486(2)  \times 10^{-2} \T \B  \\ 
0.175 &  -2.611063(2)  \times 10^{-2}& -2.3171815(9)  \times 10^{-2} &  -2.0812102(5)  \times 10^{-2} &  -1.8880110(6)  \times 10^{-2} &  -1.727142(2)  \times 10^{-2} \T \B  \\ 
0.200 &  -2.609973(3)  \times 10^{-2} & -2.316378(2)  \times 10^{-2} &  -2.080592(2)  \times 10^{-2} &  -1.887523(2)  \times 10^{-2} &  -1.726748(1)  \times 10^{-2} \T \B  \\ 
0.225 &  -2.608738(2)  \times 10^{-2}& -2.315467(3)  \times 10^{-2} &  -2.079894(5)  \times 10^{-2} &  -1.886970(2)  \times 10^{-2} &  -1.7263003(9)  \times 10^{-2}\T \B  \\ 
0.250 &  -2.607352(2)  \times 10^{-2}& -2.314445(3)  \times 10^{-2} &  -2.079112(2)  \times 10^{-2} &  -1.886353(2)  \times 10^{-2} &  -1.725800(1)  \times 10^{-2} \T \B  \\ 
 \hline\hline
 \multicolumn{1}{l}{$e$} & \multicolumn{1}{c}{$p=60$}   & \multicolumn{1}{c}{$p=65$}   & \multicolumn{1}{c}{$p =70\quad$} & \multicolumn{1}{c}{$p=75$} & \multicolumn{1}{c}{$p=80$} \T\B\\%
 \hline
0.000 &  2.6329573(1)  \times 10^{-4} &  2.2533310(3)  \times 10^{-4}& 1.950169674(6)  \times 10^{-4} &  1.7042571(5)  \times 10^{-4} &  1.5020480(2)  \times 10^{-4} \T \B  \\ %
0.050 &  -1.5922111(9)  \times 10^{-2} &  -1.475776(2)  \times 10^{-2}& -1.375078(2)  \times 10^{-2} &  -1.28716(1)  \times 10^{-2} &  -1.209737(7)  \times 10^{-2}  \T \B  \\ 
0.075 &  -1.5920996(3)  \times 10^{-2} &  -1.475686(2)  \times 10^{-2}& -1.375005(2)  \times 10^{-2} &  -1.287095(2)  \times 10^{-2} &  -1.209690(8)  \times 10^{-2} \T \B  \\ %
0.100 &  -1.591948(2)  \times 10^{-2} &  -1.475553(2)  \times 10^{-2} & -1.374891(4)  \times 10^{-2} &  -1.2869985(9)  \times 10^{-2} &  -1.2096006(4)  \times 10^{-2} \T \B  \\ %
0.125 &  -1.591752(1)  \times 10^{-2} &  -1.4753912(8)  \times 10^{-2}& -1.3747508(6)  \times 10^{-2} &  -1.2868785(9)  \times 10^{-2} &  -1.209502(1)  \times 10^{-2} \T \B  \\ %
0.150 &  -1.591513(2)  \times 10^{-2} &  -1.4751887(9)  \times 10^{-2}& -1.374583(3)  \times 10^{-2} &  -1.286731(1)  \times 10^{-2} &  -1.209375(3)  \times 10^{-2} \T \B  \\ %
0.175 &  -1.5912302(2)  \times 10^{-2} &  -1.4749509(6)  \times 10^{-2}& -1.3743806(6)  \times 10^{-2} &  -1.2865597(2)  \times 10^{-2} &  -1.209224(3)  \times 10^{-2} \T \B  \\ %
0.200 &  -1.590903(3)  \times 10^{-2} &  -1.474681(1)  \times 10^{-2}& -1.374145(5)  \times 10^{-2} &  -1.286350(4)  \times 10^{-2} &  -1.209044(5)  \times 10^{-2} \T \B  \\ %
0.225 &  -1.590534(3)  \times 10^{-2} &  -1.4743701(5)  \times 10^{-2} & -1.373887(3)  \times 10^{-2} &  -1.286132(2)  \times 10^{-2} &  -1.208854(7)  \times 10^{-2} \T \B  \\ %
0.250 &  -1.590122(2)  \times 10^{-2} &  -1.474022(2)  \times 10^{-2} & -1.373589(2)  \times 10^{-2} &  -1.285874(2)  \times 10^{-2} &  -1.208622(2)  \times 10^{-2} \T \B  \\ %
 \hline\hline
 \multicolumn{1}{l}{$e$} & \multicolumn{1}{c}{$p=85$}  & \multicolumn{1}{c}{$p=90$} &\multicolumn{1}{c}{$p=95$} & \multicolumn{1}{c}{$p =100\quad$} \T \B  \\ 
 \hline
 0.000 &  1.33377732(1)  \times 10^{-4} &  1.1922590(3)  \times 10^{-4} &  1.0721120(2)  \times 10^{-4}  & 9.692429(3)  \times 10^{-5} \T \B  \\ 
 0.050 &  -1.141067(2)  \times 10^{-2} &  -1.079745(9)  \times 10^{-2} &  -1.02464(1)  \times 10^{-2}& -9.7486(2)  \times 10^{-3}\T \B  \\ 
0.075 &  -1.141016(2)  \times 10^{-2} &  -1.0796893(5)  \times 10^{-2} &  -1.024593(2)  \times 10^{-2}  & -9.7482(3)  \times 10^{-3} \T \B  \\ 
0.100 &  -1.140936(2)  \times 10^{-2} &  -1.079628(9)  \times 10^{-2} &  -1.024536(2)  \times 10^{-2}  & -9.74776(8)  \times 10^{-3}\T \B  \\ 
0.125 &  -1.140856(2)  \times 10^{-2} &  -1.079545(8)  \times 10^{-2} &  -1.024466(7)  \times 10^{-2} & -9.7472(1)  \times 10^{-3} \T \B  \\ 
0.150 &  -1.1407409(8)  \times 10^{-2} &  -1.079445(6)  \times 10^{-2} &  -1.024384(2)  \times 10^{-2}  & -9.7464(2)  \times 10^{-3} \T \B  \\ 
0.175 &  -1.140611(1)  \times 10^{-2} &  -1.079322(3)  \times 10^{-2} &  -1.024263(3)  \times 10^{-2} & -9.74544(3)  \times 10^{-3}\T \B  \\ 
0.200 &  -1.140452(2)  \times 10^{-2} &  -1.079193(2)  \times 10^{-2} &  -1.024156(5)  \times 10^{-2} & -9.74433(6)  \times 10^{-3}\T \B  \\ 
0.225 &  -1.140282(4)  \times 10^{-2} &  -1.079042(4)  \times 10^{-2} &  -1.024012(2)  \times 10^{-2} & -9.74322(6)  \times 10^{-3}\T \B  \\ 
0.250 &  -1.140082(4)  \times 10^{-2} &  -1.078859(2)  \times 10^{-2} &  -1.023864(9)  \times 10^{-2} & -9.74155(9)  \times 10^{-3} \T \B   
\label{tbl:data}
\end{longtable*}
\end{widetext}
\vsp
\subsection{Analysis and comparisons with pN results}\label{Sec:O_e2_part_comparison} \vspace{0.1cm}

Here we set $q=1$ for notational convenience, without loss of generality, recalling that $\Delta \psi$ is linear in $q$.

\begin{figure}[h!]
 \includegraphics[width=0.9\textwidth]{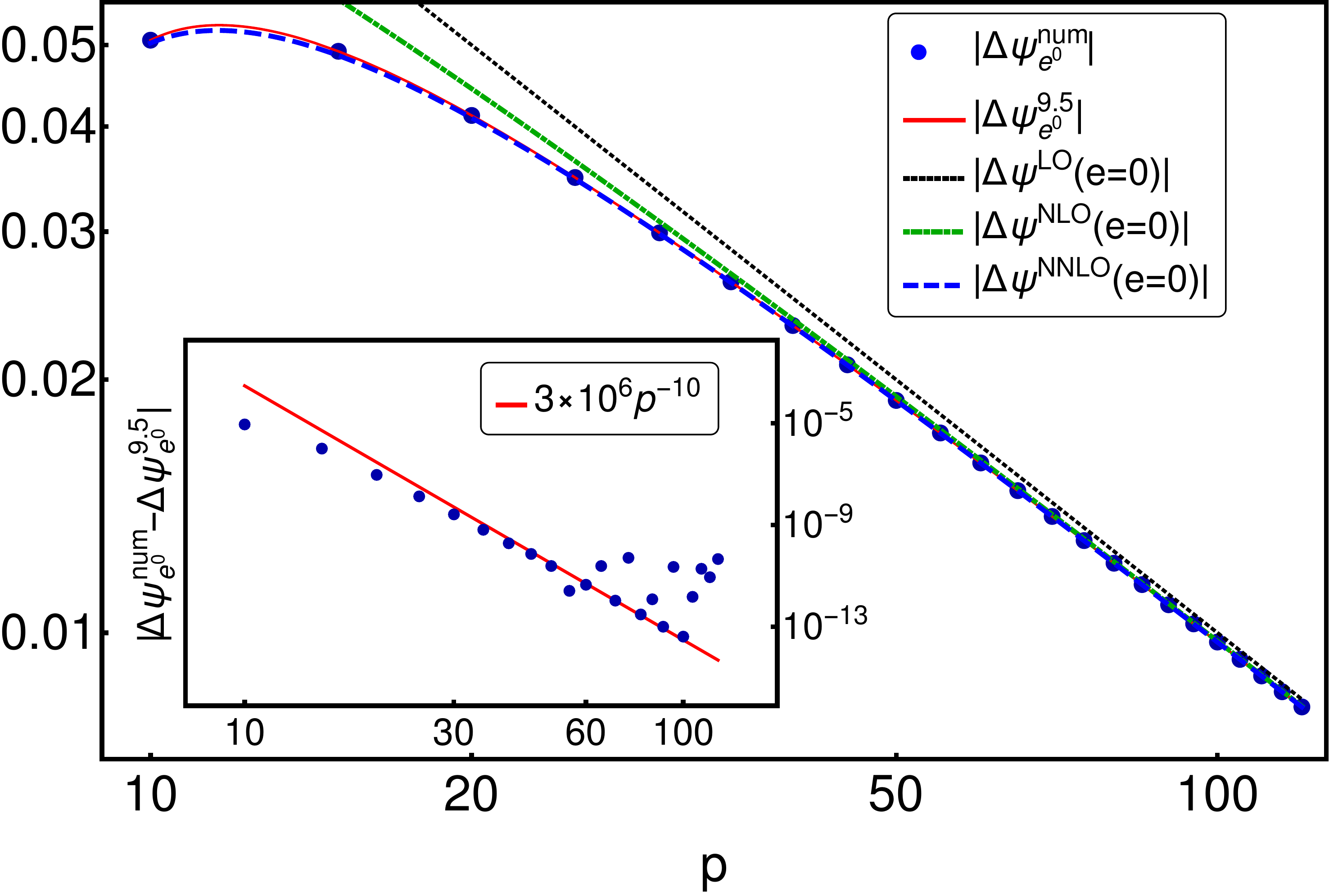}
 \caption{Comparison of the numerical values $\Delta\psi_{e^0}^\text{num}$ (blue dots) with the analytic $\Delta\psi_{e^0}^{9.5} $ (solid red curve). The difference is plotted in the inset (blue dots).  The difference appears to scale as $\ord(p^{-10})$, as shown by the red (solid) reference line in the inset. Numerical noise is clear for $p \gtrsim 50$. Both $\Delta\psi_{e^0}^\text{num}$ and $\Delta\psi_{e^0}^{9.5}$ are obtained from Eq.~(\ref{eq:circoffset}) using the results from Ref.~\cite{Akcay:2015pjz} for the former and Refs.~\cite{Shah:2015nva, Bini:2016qtx} for the latter. $\Delta\psi_{e^0}^{9.5}$ is explicitly shown in Eq.~(\ref{eq:A1}). 
The $ e= 0 $ values of our eccentric post-Newtonian expressions $\Delta\psi^{LO}, \Delta\psi^{NLO},\Delta\psi^{NNLO}$ are displayed in the main plot as the dotted black, dot-dashed green and dashed blue lines, respectively. As expected, these approximate the numerical data better with increasing pN order and increasing $p$.
We generated additional $\Delta\psi_{e^0}^\text{num}$ data for this plot that is not provided in Table \ref{table:Numerical_Data}: for $p = \{105,110,115,120\}$, $\Delta\psi_{e^0}^\text{num} = \{ 8.8049204(2)  \times 10^{-5},8.0339083(2)  \times 10^{-5} ,  7.359865(2)  \times 10^{-5} ,  6.7671891(9)  \times 10^{-5}\} $.
 }
 \label{Fig:e20_limit_plot}
\end{figure}

First, consider the limit of zero eccentricity. Figure \ref{Fig:e20_limit_plot} compares numerical results for $\Delta\psi_{e^0}^\text{num} \equiv \Lim{e\to 0}\Delta\psi$ (blue dots) with the analytic post-Newtonian expansion $\Delta\psi_{e^0}^{9.5}$ (red solid line). The numerical results for $\Delta\psi_{e^0}^\text{num}$ were obtained via Eq.~(\ref{eq:Delta_psi_e20_limit}) and (\ref{eq:circoffset}), with the correction term $\Delta\Phi^{e\to0}$ determined by the EOB $\rho$ function, with $\rho$ constituted from the circular-orbit redshift invariant $z$ and the EOB $\bar{d}$ function, and with the latter numerically computed to high accuracy in Ref.~\cite{Akcay:2015pjz}. The analytic result $\Delta\psi_{e^0}^{9.5}$ was described in Sec.~\ref{sec:circlimit} and is given explicitly in Eq.~(\ref{eq:A1}) of Appendix A. 
The numerical and analytical results are in robust agreement, as indicated by the fact that the (red) curve passes through all the numerical data points in Fig.~\ref{Fig:e20_limit_plot} (noting that the numerical error bars are obscured by the points themselves). The difference $|\Delta\psi_{e^0}^\text{num} - \Delta\psi^{9.5}_{e^0}|$ is shown in the inset (blue dots), and is compared against a reference line $\alpha p^{-10}$ with $\alpha = 3\times 10^6$. The inset provides evidence that the difference scales as $p^{-10}$ at large $p$, indicating that the residual disagreement is principally due to the truncation of the post-Newtonian series, which is known only up to $\ord(p^{-19/2})$. 
The scattered nature of the dots in the inset is due to noise in the numerically-computed $\Delta\psi^\text{circ}$. 

The dotted and dashed lines on Fig.~\ref{Fig:e20_limit_plot} indicate the eccentric-orbit post-Newtonian result (\ref{eq:Delta_psi_p_e}) evaluated at $e=0$ at LO (dotted black), NLO (dot-dashed green) and NNLO (dashed blue). (N.B.~The latter expressions agree with $ \Delta\psi^{9.5}_{e^0}$ at the relevant orders.) As expected, these lines provide successively-better approximations to the numerical results at large $p$.

\begin{figure}[h]
 \includegraphics[height=0.85\textheight]{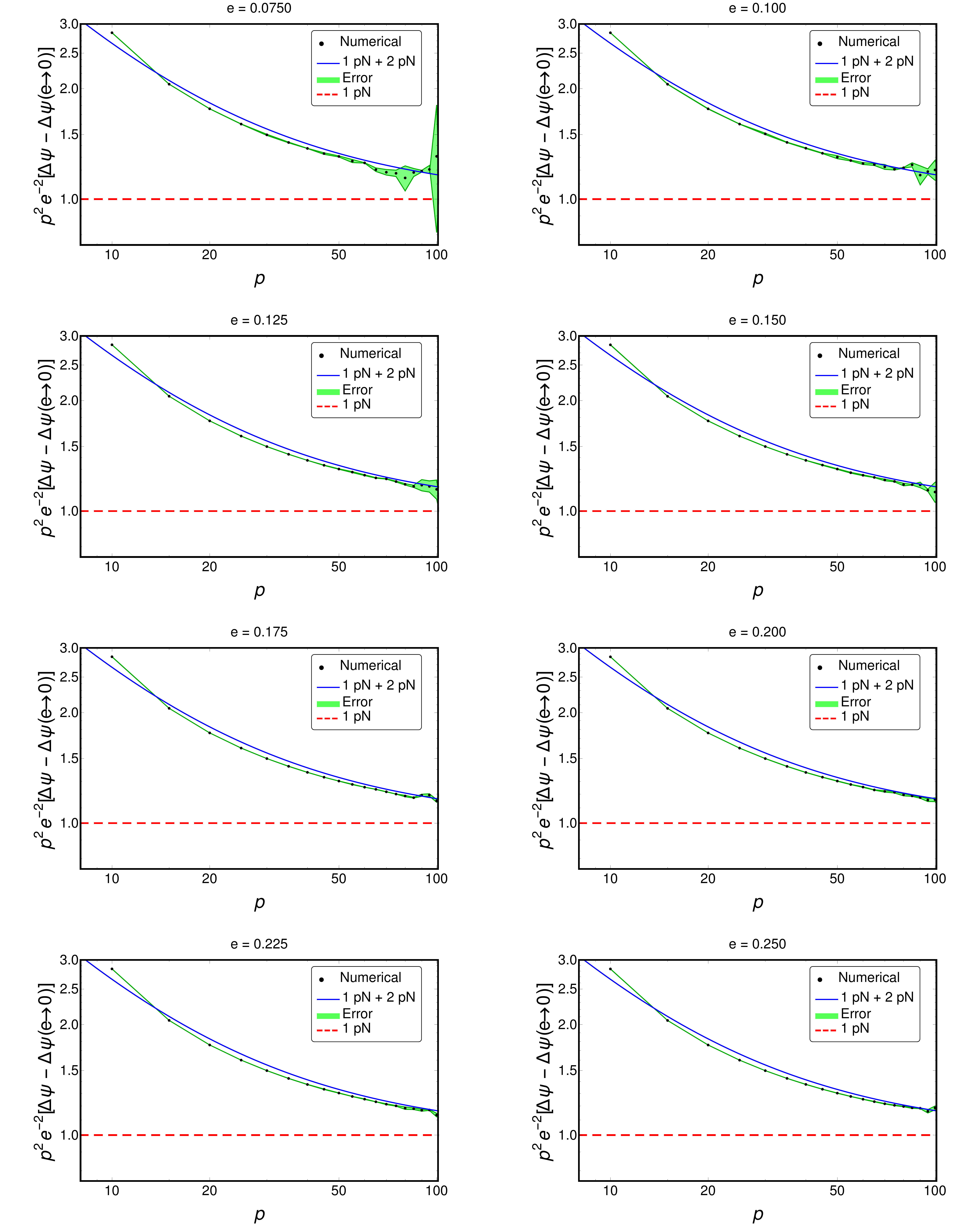}
 \caption{The numerical residual $ e^{-2}p^2(\Delta\psi- \Delta \psi_{e^0}^\text{num}) $ represented by the black dots for $e=\{ 0.075, 0.1, 0.125, \ldots, 0.225, 0.25 \} $. This is compared with $e^{-2}p^2\Delta\psi_{e^2}\approx 1 + \f{16.5705}{p}  $ shown by the solid blue curves labelled as ``1 pN + 2 pN'' in the panels. The numerical residuals  contain contributions from the unknown $\ord\left( p^{\le -4}\,e^{\ge 4}\right) $ terms hence the offset between the solid blue curves and the black dots, which decreases with increasing $p$ as expected. The red dashed lines mark the 1 pN term $e^{-2}p^2\times(e^2/p^2)=1$. The green strips are our estimated numerical errors. 
 Although we calculate the errors only at the $p$ values given in Table~\ref{tbl:data} the error bars in the plots are connected with green shaded curves for visualization purposes. The data is noisier for low $e$ and high $p$ values where the magnitude of the term that we extract from the numerical data becomes comparable in size to the estimated errors. In the axis labels, $\Delta \psi_{e^0}^\text{num}$ is denoted by $\Delta\psi(e\to 0)$.}
\label{Fig:1PN_2PN_plot}
\end{figure}

Next we confront our numerical data for $\Delta\psi^\text{num} $ with the eccentric-orbit post-Newtonian result at NNLO, Eq.~(\ref{eq:Delta_psi_p_e}). If the results are compatible, we expect to find
\be
\Delta\psi^\text{num} =  \Delta\psi_{e^0}^\text{num} + \Delta\psi_{e^2} + \Delta\psi_{e^4} + \ord(e^{\ge2}p^{\le-4})\label{eq:Delta_psi_e20_eq} ,
\ee
where
\begin{align}
\Delta\psi_{e^2} &= \left[1+ \left(\f{341}{16}-\f{123\pi^2}{256}\right) \f{1}{p}\right]\f{e^2}{p^2}, 
\qquad \qquad \Delta\psi_{e^4} = -\f{e^4}{2p^3} .
\label{eq:Delta_psi_e_squared_term}
\end{align}
%
Figure \ref{Fig:1PN_2PN_plot} shows numerical data for $e^{-2}p^2(\Delta\psi^\text{num}-\Delta\psi_{e^0}^\text{num}) $ compared against $e^{-2}p^2 \Delta\psi_{e^2} \approx 1 + 16.5705/p$, for several values of eccentricity  $e=\{ 0.075, 0.1, 0.125, \ldots, 0.25 \}$. 
The plots show that the numerical data (black dots within shaded green confidence limits) is in robust agreement with the pN series truncated at NNLO (blue lines). As expected for a pN series, the offset between the blue curves and the numerical data decreases as $p$ increases. The residual disagreement is likely to arise from the as-yet-unknown higher-order pN terms at orders $\ord(e^2 p^{-4}), \ord(e^2 p^{-5})$, as well as from the known term at $\ord(e^4 p^{-3})$. Our data set becomes noisy for increasing values of $p$ and smaller values of $e$, as indicated by the upper half of the plots in Fig.~\ref{Fig:1PN_2PN_plot}. In this regime, the magnitude of 
$16.5705 \,e^2/p^3$ is sufficiently small that it is comparable to the numerical error itself.

In principle, numerical data can be used to constrain the (as-yet-unknown) higher-order coefficients of the eccentric-orbit pN series. For instance, the fact that the curves cross over in Fig.~\ref{Fig:1PN_2PN_plot} (cf.~the blue line and green band) hints at a change of sign in the coefficients of the pN series. We obtained numerical estimates for the values of the known $\ord(e^2 p^{-3})$ ($\approx 16.5705$) and the unknown $\ord(e^2 p^{-4})$ and $\ord(e^2 p^{-5})$ pN coefficients by fitting $ p^3 e^{-2}(\Delta\psi^\text{num}-\Delta\psi_{e^0}^\text{num}-e^2 p^{-2})$ for a range of $\{p,e\}$ to a model of the form $b_2 + b_3/p + b_4/p^2 + \ldots$, using only the `cleanest' portion of our data, i.e., $p\le 45$. We varied the number of fit parameters from two to five and obtained the best fits using standard $\chi^2$-minimizing techniques. Our best-fit results were
\begin{align}
b_2 &= \ \ 15.95 \pm 0.45 \ \ \text{  vs. } 16.5705\ldots, \nn \\
b_3 &= -55.3 \pm 13.1, \ \ \text{and  } \ b_4 =  793 \pm 86 \label{eq:best_fit_params},
\end{align}
with errors quoted at the $3\sigma$ level.
The estimate for $b_2$ is compatible with our analytical 2pN result, and $b_3$ has the opposite sign to $b_2$ as expected from the crossing of the green and blue curves in Fig.~\ref{Fig:1PN_2PN_plot}.
Here we used simple fitting functions, omitting the $\ln(p)$ terms starting at $\ord(p^{-5})$ that are too small for our code to constrain at its current level of accuracy. 
It is likely that in this range, $p \le 45$, the higher-order unknown pN terms are large enough to `contaminate' our
estimates for $b_{\ge3}$. Consequently, the errors bars quoted above may well prove to be underestimates. 
Since the data with the smallest errors (the largest statistical weight) is located at $p\le 20$, where sign changes may occur with each new term added to the pN series, we may not even be fully confident of the signs of $b_3, b_4$. Nonetheless, we have presented our best estimates here with future work in mind.

\begin{figure}[ht]
 \includegraphics[width=0.9\textwidth]{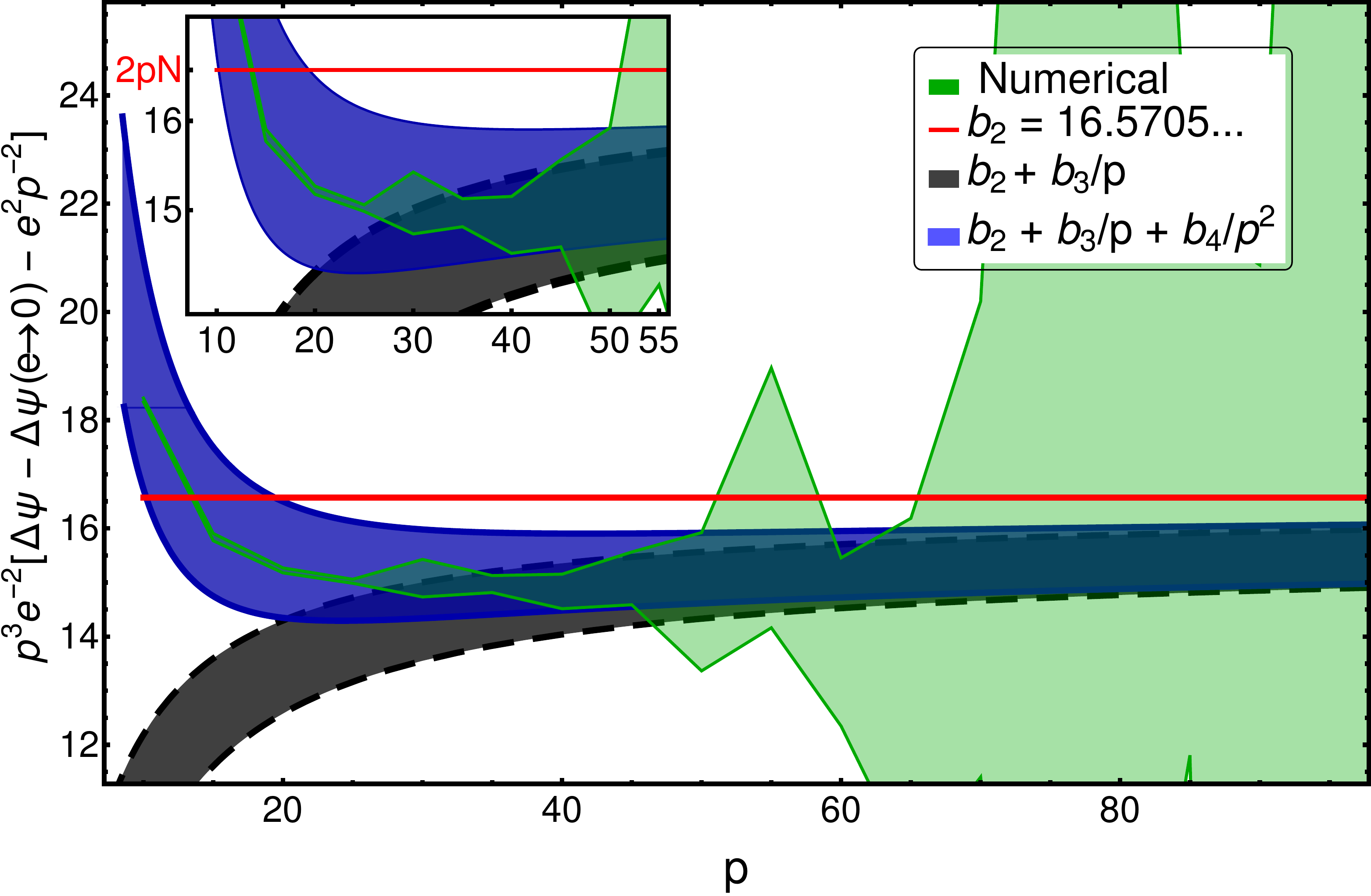}
\caption{Numerical extraction of the subdominant $\ord(e^2/p^3), \ord(e^2/p^4),  \ord(e^2/p^5)$ terms from our data. Our pN derivation yields $b_2\equiv  \left({341}/{16}-{123\pi^2}/{256} \right) \approx 16.5705$
for the coefficient of the $\ord(e^2/p^3)$ term which we plot as the thick red line. The numerical residual $p^3e^{-2}[\Delta\psi - \Delta \psi_{e^0}^\text{num} - e^2p^{-2}\, ]$ is represented by the green shaded error region. Since our
error grows with increasing $p$ our data is only `clean' for $10\le p \lesssim 45 $ where we can estimate the coefficients of the $ \ord(e^2/p^4),  \ord(e^2/p^5)$ terms. The black region between the dashed, black curves represents our numerical estimation $b_2 +b_3/p$ with $b_3 = -55.3\pm 13.1$. The blue region between the solid, blue curves is our numerical estimation $b_2 +b_3/p+b_4/p^2$ with $b_4 = 793\pm86$. See Sec.~\ref{Sec:O_e2_part_comparison} for the details of these parameter extractions.
The error region for our numerical residual grows rather large for $p\gtrsim 45$ mostly because of our low-eccentricity data where the absolute magnitude of the residual becomes comparable to our error.
}
\label{Fig:2PN_plot}
\end{figure}

Figure \ref{Fig:2PN_plot} illustrates the fitting of higher-order terms to the numerical data. The green shaded region indicates confidence limits for the quantity $p^3e^{-2}(\Delta\psi^\text{num} - \Delta\psi_{e^0}^\text{num}-e^2 p^{-2})$, numerically computed from data for $\Delta\psi^\text{num}$. The straight red line marks the analytically known $\ord(e^2\,p^{-3}) $ pN term, which is approximately equal to $16.5705$. The black region (bounded by two dashed curves) shows the pN series up to and including the $\ord(e^2\, p^{-4})$ term with coefficient $b_3$ [Eq. (\ref{eq:best_fit_params})], and the blue region (bounded by solid curves) shows the pN series up to and including the $\ord(e^2\, p^{-5})$ term with coefficient $b_4$ [Eq.~(\ref{eq:best_fit_params})]. We note that the latter is consistent with the numerical data across the whole range in $p$.

To improve the estimates of the pN parameters, one would need to perform the numerical extractions at large $p$ and small $e$, ideally $p>1000$ and $e<10^{-3}$. As the plots show, our data is noisy in the large $p$ regime, and the errors become comparable in the magnitude to the pN terms themselves. This is not altogether surprising, as the Lorenz-gauge code is unsuited to weak-field, small-eccentricity applications; whereas (forthcoming) radiation-gauge codes may probe this regime effectively (Sec.~\ref{Sec:discussion}). Furthermore, we cannot, at present, estimate the coefficient of the $\ord(e^4/p^3)$ term with any confidence. In our current range of eccentricities, $e \le 0.25$, the magnitude of this term is comparable to our estimated error for
$\Delta\psi^\text{num}$. A reasonable numerical estimation of this term requires several improvements which we explain in detail in Sec.~\ref{Sec:discussion}. 

Despite such caveats, we now proceed to synthesize our numerically-acquired information with the analytical knowledge from the pN series. Extracting $\ord(p^{-4})$, $\ord(p^{-5})$ parts of $\Delta\psi_{e^0}^{9.5}$ from Eq.~(\ref{eq:A1}) and combining these with our estimates above leads to the following expression: 
\begin{align}
q^{-1}  \Delta{\psi} = &- p^{-1}+\left(\f{9}{4}+e^2\right) p^{-2} +\left[\f{739}{16}-\f{123\pi^2}{64}+\left(\f{341}{16}-\f{123\pi^2}{256}\right)e^2-\f{e^4}{2}\right] p^{-3}\nn\\  &+ \left[a_4+a_4^\text{log}\ln p -55(13)\, e^2 + \ord(e^4) \right] p^{-4}+\left[a_5 + a_5^\text{log}\ln p +793(86)e^2+\ord(e^4)\right] p^{-5} + \ord(p^{-6})\label{eq:Delta_psi_p_e_approx} ,
\end{align}
where
\begin{align}
a_4 &= \frac{31697 \pi ^2}{6144}+\frac{1256 \gamma }{15}-\frac{587831}{2880}+\frac{729 \ln (3)}{5}+\frac{296 \ln (2)}{15} \approx 68.997, \label{eq:a4} \\
a_5 &= \frac{2479221 \pi ^2}{8192}-\frac{22306 \gamma }{35}-\frac{48221551}{19200}-\frac{31347 \ln (3)}{28}+\frac{22058 \ln (2)}{105}\approx -976.799, \label{eq:a5} \\
a_4^\text{log} &= -\frac{628}{15} \approx  41.867,\qquad a_5^\text{log} =\frac{11153}{35} \approx 318.657.\label{eq:a4Log_a5Log}
\end{align}

\vsp
\section{Discussion}\label{Sec:discussion} \vspace{0.1cm}

In the preceding sections we have computed the spin precession scalar $\psi$ for eccentric compact binaries via two complementary approaches. In the GSF approach, we obtained $\psi$ in the strong-field regime at $\ord(q)$, whereas in the pN approach we obtained $\psi$ at arbitrary mass ratio $q$ as an expansion in powers of $c^{-2}$ in the weak-field. We have established here that the results are in agreement at (separately) $\ord(q^0)$; at $\ord(q^1)$ in the circular limit; and at $\ord(q e^2 p^{-2})$ and $\ord(q e^2 p^{-3})$ for eccentric orbits. 
We also obtained evidence for a sign change at $\ord(q e^2 p^{-4})$, as well as for the likely sign of the term at $\ord(q e^2 p^{-5})$, with the caveat that the error bars on these quantities are large in magnitude (see Eq.~(\ref{eq:best_fit_params})). The results of our code also agree with the analytical log and non-log $\ord(q e^0)$ terms up to $\ord(p^{-19/2})$, as illustrated by the inset of Fig.~\ref{Fig:e20_limit_plot}.

%

To overcome certain limitations of our Lorenz-gauge numerical implementation --- such as its insufficient relative accuracy at large $p$, highlighted in Fig.~\ref{Fig:2PN_plot} --- we propose that $\Delta \psi$ should now be calculated via further complementary approaches. One possibility is to apply the \emph{radiation-gauge} GSF architecture to compute $\psi$ for eccentric orbits at much greater numerical precision \cite{vandeMeent:2015lxa, vandeMeent:2016pee}. This approach may allow one to compute $\Delta\psi$ close to the separatrix ($p=6+2e$) of bound orbits in Schwarzschild spacetime, to quantify the (anticipated) divergence of $\Delta\psi$ in this limit. Another possibility is to use S.~Hopper's doubly-expanded (in $p,e$) expressions for $h_{ab}$ and its derivatives to obtain a pN expression for $\Delta\psi$ accurate to $\ord(e^{10})$ and $\ord(p^{-5})$ \cite{Hopper:2015icj, Hopper:pvt}. A third possibility is to extend the approach of Bini, Damour \& collaborators \cite{Bini:2014zxa, Bini:2015kja, 
Bini:2015bla, Kavanagh:2015lva} which makes expert use of the Mano-Suzuki-Takasugi formalism \cite{Mano:1996vt}. 

Extending the \emph{arbitrary mass ratio} pN calculation of $\psi$ to next order (NNNLO) presents a stiff challenge. 
The NNNLO spin-orbit Hamiltonian has not appeared in the literature. In principle, one can obtain it from the 4pN metric for non-spinning binaries\footnote{We thank Thibault Damour for pointing this out.}.
This computation requires the attention of experts of post-Newtonian theory.

The boundary between GSF and numerical relativity is under active exploration. Recently, the redshift invariant $z$ was extracted from numerical relativity simulations of quasi-circular black hole binaries, via the helical (quasi-)Killing vector field \cite{Zimmerman:2016ajr}. It is plausible that the circular-orbit precession invariant $\psi^{\text{circ}}$ can be extracted from the derivatives of the helical Killing vector field (see Eq.~(3) in Ref.~\cite{Dolan:2013roa}). Obtaining the eccentric-orbit precession invariant poses a stiffer challenge as, in the absence of a continuous symmetry, it is necessary to track the spin of the small black hole in some sufficiently coordinate-insensitive way.

Our prescription for computing the spin precession invariant $\Delta \psi$ is, at present, far less elegant than that available for computing the redshift invariant $\Delta z$. Remarkably, in Ref.~\cite{Akcay:2015pza} it was shown that $\Delta z = - \left<H \right> z$, where $\left< H \right> = \frac{1}{\mathcal{T}} \int_{0}^\mathcal{T} \frac{1}{2} h_{00} d\tau$. This offered a significant simplification of the earlier prescription of Ref.~\cite{Barack:2011ed}, leading to improved numerical accuracy and physical insight. It is natural to speculate as to whether a similar simplification may be possible for the spin-precession calculation. To consider this, let us recap the argument for $z$. First, following the approach of Sec.~\ref{Sec:Outline_GSF_method}, and using Eqs.~(\ref{eq:energy}), (\ref{eq:delta_Y_generic}) and (\ref{eq:ELrdot}), one may quickly establish that $\delta \mathcal{T} = E \delta T - L \delta \Phi - \int_0^{\mathcal{T}} \frac{1}{2} h_{00} d \tau$. Second, it follows from the definition of 
$\Delta$ that $\Delta z = \Delta \left( \mathcal{T} / T \right) = \Delta \mathcal{T} / T$ (with $\Delta T = 0$ by definition). Third, combining Eq.~(\ref{eq:Delfromdel}) with the first step yields
\begin{eqnarray}
\Delta \mathcal{T} &=& \delta \mathcal{T} - \frac{\partial \mathcal{T}}{\partial T} \delta T - \frac{\partial \mathcal{T}}{\partial \Phi} \delta \Phi , \\
 &=& - \int_0^{\mathcal{T}} \frac{1}{2} h_{00} d \tau + \left(E - \frac{\partial \mathcal{T}}{\partial T} \right)  \delta T  + \left( -L - \frac{\partial \mathcal{T}}{\partial \Phi} \right) \delta \Phi.
\end{eqnarray}
Finally, it may be shown using the first law of binary mechanics \cite{Tiec:2013kua} that the two bracketed terms are identically zero (see Appendix B of \cite{Akcay:2015pza}), and thus the simple result for $\Delta z$ follows. Thus, it seems that we lack two crucial ingredients to transfer the recipe for $\Delta z$ to $\Delta \psi$. First, an appropriate analogue of the energy equation (\ref{eq:energy}) involving $\Psi$, and second, an appropriate analogue of the first law. It is possible that the laws of binary mechanics for spinning bodies \cite{Blanchet:2012at} can provide the missing insight here.

In summary, we have taken one more small step in extracting physical content from GSF theory, to move further towards the goal of accurately modelling the gravitational two-body problem. There remain many challenges ahead: from transcribing eccentric-orbit GSF results into EOB theory, to calculating invariants for eccentric orbits on Kerr, and, most importantly, extending GSF theory to second order in the mass ratio. Progress is being made on a range of fronts, inspired by the successes of LIGO and eLISA Pathfinder that have heralded a new era of gravitational wave astronomy.

\vsp

\acknowledgements
\vspace{0.1cm}
S.A.~acknowledges support from the Irish Research Council, funded under the National Development Plan for Ireland. S.D.~acknowledges support under EPSRC Grant No.~EP/M025802/1, and from the Lancaster-Manchester-Sheffield Consortium for Fundamental Physics under STFC Grant No.~ST/L000520/1. We are indebted to Alexandre Le Tiec for his assistance with the pN calculation (Sec.~\ref{Sec:pN_expansion}), and for discussions and correspondence relating to Sec.~\ref{subsec:geodetic} and Sec.~\ref{sec:circlimit}. We are indebted to Barry Wardell for calculating previously-unpublished regularization parameters
for $h_{ab}$ on eccentric orbits (Sec.~\ref{sec:Reg_params}). We are grateful to Thibault Damour and Donato Bini for suggestions on improving the manuscript and corrections. S.A. also thanks Chris Kavanagh, Seth Hopper, Adrian Ottewill and Niels Warburton. 

\vspace{0.3cm}

\appendix
\section{$\Delta\psi_{e^0}^{9.5} $: the pN series for $\lim_{e\to0}\Delta\psi$ up to $p^{-19/2}$}\label{sec:AppA} \vspace{0.1cm}
\noindent
Below $\gamma$ denotes the Euler's constant and $\zeta$ is the Riemann zeta function.
\begin{align}
\Delta\psi_{e^0}^{9.5} &=-\f{1}{p}+\frac{9}{4 p^2}+\frac{2956-123 \pi ^2}{64 p^3}+\frac{\frac{628}{15} \ln \left(\frac{1}{p}\right)+\frac{31697 \pi ^2}{6144}+\frac{1256 \gamma
   }{15}-\frac{587831}{2880}+\frac{729 \ln (3)}{5}+\frac{296 \ln (2)}{15}}{p^4}\nn \\
   &+\frac{-\frac{11153}{35} \ln \left(\frac{1}{p}\right)+\frac{2479221 \pi ^2}{8192}-\frac{22306 \gamma
   }{35}-\frac{48221551}{19200}-\frac{31347 \ln (3)}{28}+\frac{22058 \ln (2)}{105}}{p^5}+\frac{49969}{315} \pi  \left(\frac{1}{p}\right)^{11/2}\nn\\
   +&\frac{-\frac{344021 \ln \left(\frac{1}{p}\right)}{3780}-\frac{7335303 \pi ^4}{131072}+\frac{7230119267 \pi
   ^2}{2359296}-\frac{344021 \gamma }{1890}-\frac{1900873914203}{101606400}+\frac{9765625 \ln (5)}{9072}+\frac{234009
   \ln (3)}{70}-\frac{2514427 \ln (2)}{270}}{p^6}\nn\\
   +& \frac{-\frac{849152 \ln ^2\left(\frac{1}{p}\right)}{1575}+\left(\frac{1316474014843}{87318000}-\frac{3396608 \gamma
   }{1575}-\frac{2574848 \ln (2)}{1575}-\frac{468018 \ln (3)}{175}\right) \ln \left(\frac{1}{p}\right)+\frac{63488
   \zeta (3)}{15}+\frac{341587582057 \pi ^4}{1006632960}}{p^7}\nn\\
   +&\f{\frac{25657561505749 \pi ^2}{2477260800}-\frac{3396608 \gamma
   ^2}{1575}-\frac{1282190594044678657}{7041323520000}-\frac{468018 \ln ^2(3)}{175}-\frac{931328 \ln
   ^2(2)}{1575}+\frac{157464 \ln (6)}{55}-\frac{361328125 \ln (5)}{24192}}{p^7}\nn\\
   +&\f{-\frac{936036}{175} \ln (2) \ln
   (3)+\frac{257433623847 \ln (3)}{8624000}+\gamma  \left(\frac{1316474014843}{43659000}-\frac{5149696 \ln
   (2)}{1575}-\frac{936036 \ln (3)}{175}\right)+\frac{2658265157683 \ln (2)}{43659000}}{p^7}\nn \\
   +&\frac{\frac{44326552 \ln ^2\left(\frac{1}{p}\right)}{11025}+\left(-\frac{1884153630595993}{31783752000}+\frac{177306208
   \gamma }{11025}+\frac{1360096 \ln (2)}{11025}+\frac{59742279 \ln (3)}{2450}\right) \ln
   \left(\frac{1}{p}\right)-\frac{861696 \zeta (3)}{35}}{p^8}\nn\\
   +&\f{-\frac{623848083842333 \pi ^4}{21474836480}+\frac{569460279231731
   \pi ^2}{123312537600}+\frac{177306208 \gamma
   ^2}{11025}+\frac{78550205239878250993769}{28193459374080000}+\frac{59742279 \ln ^2(3)}{2450}}{p^8}\nn\\
   +&\f{-\frac{520925728 \ln
   ^2(2)}{11025}+\frac{678223072849 \ln (7)}{92664000}-\frac{849557646 \ln (6)}{17875}+\frac{8570767578125 \ln
   (5)}{96864768}+\frac{59742279 \ln (2) \ln (3)}{1225}}{p^8}\nn\\
   +&\f{\gamma  \left(-\frac{1888832198890393}{15891876000}+\frac{2720192 \ln (2)}{11025}+\frac{59742279 \ln
   (3)}{1225}\right)+\frac{42057788451157 \ln (2)}{5297292000}-\frac{719273429413893 \ln
   (3)}{3139136000}}{p^8}\nn\\
   +&\frac{-\frac{559060184 \ln
   ^2\left(\frac{1}{p}\right)}{218295}+\left(-\frac{471911654529055879}{4195455264000}-\frac{2236240736 \gamma
   }{218295}+\frac{592031820992 \ln (2)}{3274425}-\frac{1115809587 \ln (3)}{13475}-\frac{76708984375 \ln
   (5)}{3143448}\right) \ln \left(\frac{1}{p}\right)}{p^9}\nn\\
   &+\f{-\frac{2231776 \zeta (3)}{63}+\frac{128148402261 \pi
   ^6}{67108864}-\frac{4415389705519783271 \pi ^4}{6597069766656}+\frac{1137608424260437147 \pi
   ^2}{9766352977920}-\frac{2236240736 \gamma
   ^2}{218295}}{p^9}\nn\\
   +&\f{\frac{1311867260284968501440736791}{20734275551109120000}-\frac{76708984375 \ln
   ^2(5)}{3143448}-\frac{1115809587 \ln ^2(3)}{13475}+\frac{7359693939328 \ln ^2(2)}{9823275}-\frac{4747561509943 \ln
   (7)}{33696000}}{p^9}\nn\\
   +&\f{+\frac{88880613849 \ln (6)}{250250}-\frac{76708984375 \ln (2) \ln
   (5)}{1571724}+\frac{1091431829233203125 \ln (5)}{7249746696192}+\frac{780367277518947 \ln (3)}{1104975872}-\frac{51965633780407594519 \ln (2)}{18879548688000}}{p^9}\nn\\
   +&\f{\gamma 
   \left(-\frac{474026163982111879}{2097727632000}+\frac{1184063641984 \ln (2)}{3274425}-\frac{2231619174 \ln
   (3)}{13475}-\frac{76708984375 \ln (5)}{1571724}\right)-\frac{2231619174 \ln (2) \ln
   (3)}{13475}}{p^9}\nn\\
   &-\frac{2620819 \pi  \left(\frac{1}{p}\right)^{13/2}}{2100}+\frac{2782895449 \pi  \left(\frac{1}{p}\right)^{15/2}}{2910600}
   +\left(\frac{1}{p}\right)^{17/2} \left(-\frac{588455702 \pi  \ln \left(\frac{1}{p}\right)}{165375}+\frac{10999172 \pi
   ^3}{4725}\right)	\nn\\
   &+\left(\frac{1}{p}\right)^{17/2}\pi  \left(\frac{106497918450629063}{2097727632000}-\frac{1176911404 \gamma }{165375}-\frac{1001054764 \ln
   (2)}{165375}-\frac{50077926 \ln (3)}{6125}\right)\nn\\
   +&\left(\frac{1}{p}\right)^{19/2} \left(\frac{23055449891 \pi  \ln \left(\frac{1}{p}\right)}{771750}-\frac{180728953 \pi
   ^3}{11025}\right)\label{eq:A1}\\
   +&\left(\frac{1}{p}\right)^{19/2}\pi  \left(-\frac{3130119243444996194647}{11453592870720000}+\frac{23055449891 \gamma
   }{385875}+\frac{316521883 \ln (2)}{15435}+\frac{6854694417 \ln (3)}{85750}\right)\nn
\end{align}

\bibliographystyle{apsrev4-1}
\bibliography{references}

\end{document}